\documentclass[lettersize,journal]{IEEEtran}
\usepackage{amsmath,amsfonts}
\usepackage{algorithmic}
\usepackage{algorithm}
\usepackage{array}
\usepackage[caption=false,font=normalsize,labelfont=sf,textfont=sf]{subfig}
\usepackage{textcomp}
\usepackage{stfloats}
\usepackage{url}
\usepackage{verbatim}
\usepackage{graphicx}
\usepackage{cite}
\usepackage{xcolor}
\usepackage{nomencl}
\usepackage{etoolbox}
\usepackage{caption} 
\usepackage{placeins}
\usepackage{booktabs}  
\usepackage{multirow}  
\usepackage{rotating}  
\usepackage{tabularx}  

\makenomenclature

\renewcommand{\nomgroup}[1]{%
  \ifthenelse{\equal{#1}{A}}{\item[\textbf{Sets and Indices}]}{}%
  \ifthenelse{\equal{#1}{B}}{\item[\textbf{Parameters}]}{}%
  \ifthenelse{\equal{#1}{C}}{\item[\textbf{Variables}]}{}%
}
\title{Generation Expansion Planning with Upstream Supply Chain Constraints on Materials, Manufacturing, and Deployment}

\author{Boyu~Yao, Andrey~Bernstein, Yury~Dvorkin}

\begin{document}

\maketitle

\begin{abstract}
Rising electricity demand underscores the need for secure and reliable generation expansion planning that accounts for upstream supply chain constraints. Traditional models often overlook limitations in materials, manufacturing capacity, lead times for deployment, and field availability, which can delay availability of planned resources and thus to threaten system reliability. This paper introduces a multi-stage supply chain-constrained generation expansion planning (SC-GEP) model that optimizes long-term investments while capturing material availability, production limits, spatial and temporal constraints, and material reuse from retired assets. A decomposition algorithm efficiently solves the resulting MILP. A Maryland case study shows that supply chain constraints shift technology choices, amplify deployment delays caused by lead times, and prompt earlier investment in shorter lead-time, low-material-intensity options. In the low-demand scenario, supply chain constraints raise investment costs by \$1.2 billion. Under high demand, persistent generation and reserve shortfalls emerge, underscoring the need to integrate upstream constraints into long-term planning.
\end{abstract}


\IEEEpeerreviewmaketitle

\printnomenclature

\nomenclature[A01]{$\mathcal{I},\mathcal{J}$}{Set of zones; indexed by $i, j$.}%
\nomenclature[A02]{$\mathcal{L}/\mathcal{LS}_{l}/\mathcal{LR}_{l}$}{Set of transmission corridors and subsets for sending/receiving in zone $i$; indexed by $l$.}%
\nomenclature[A03]{$\mathcal{K}/\mathcal{N}$}{Set of technologies (\textit{spv} – solar PV, \textit{lbw} – land-based wind, \textit{osw} – offshore wind, \textit{bse} – battery storage); indexed by $k$. Set of types (\textit{th} – thermal, \textit{rn} – renewable, \textit{st} – storage); indexed by $n$.}%
\nomenclature[A04]{$\mathcal{G}/\mathcal{G}_{i}/\mathcal{G}^{n}/\mathcal{G}^{k}/\bar{\mathcal{G}}/\tilde{\mathcal{G}}$}{Set of generators and storage units, including subsets by zone $i$, type $n$, technology $k$, existing units ($\bar{\mathcal{G}}$), and candidate units ($\tilde{\mathcal{G}}$); indexed by $g$.}%
\nomenclature[A05]{$\mathcal{M}$}{Set of critical materials; indexed by $m$.}%
\nomenclature[A06]{$\mathcal{C}$}{Set of components; indexed by $c$.}%
\nomenclature[A07]{$\mathcal{P}/\mathcal{P}^{k}$}{Set of products and subsets by technology $k$; indexed by $p$.}%
\nomenclature[A08]{$\mathcal{Y}$}{Set of years; indexed by $y$.}%
\nomenclature[A09]{$\mathcal{T}$}{Set of representative days per year $y$; indexed by $t$.}%
\nomenclature[A10]{$\mathcal{H}$}{Set of hours per day $t$; indexed by $h$.}%

\nomenclature[B01]{$L_{ithy}$}{Load demand in zone $i$ day $t$ hour $h$ year $y$ (MW).}%
\nomenclature[B02]{$\overline{L}_{y}$}{System peak load demand in year $y$ (MW).}%
\nomenclature[B03]{$\overline{P}^{\text{G}}_g/\overline{P}^{\text{L}}_l$}{Power capacity of unit $g$ and transfer capacity of transmission corridor $l$ (MW).}%
\nomenclature[B04]{$F^{\text{GEN}}_{igthy}$}{Availability factor of renewable generation unit $g$ in day $t$ hour $h$ year $y$ and zone $i$ (unitless).}%
\nomenclature[B05]{$\overline{E}^{\text{ST}}_g$}{Energy capacity of storage unit $g \in \mathcal{G}^{\text{st}}$ (MWh).}%
\nomenclature[B06]{$\epsilon^{\text{CH}}/\epsilon^{\text{DC}}$}{Charging/discharging efficiencies of storage unit $g \in \mathcal{G}^{\text{st}}$ (unitless).}%
\nomenclature[B07]{$F^{\text{ELCC}}_{ky}$}{ELCC factor of technology $k$ in year $y$ (unitless).}%
\nomenclature[B08]{$R^{\text{RM}}_{y}$}{System-wide reserve margin in year $y$ (unitless).}%
\nomenclature[B09]{$R^{\text{RPS}}_{ky}$}{Renewable portfolio standard (RPS) mandate for technology $k$ in year $y$ (unitless).}%
\nomenclature[B10]{$P^{\text{RM}}$}{Penalty cost for reserve margin violation (\$/MW).}
\nomenclature[B10]{$P^{\text{RPS}}$}{Penalty cost for RPS non-compliance (\$/MWh).}
\nomenclature[B11]{$P^{\text{VOLL}}$}{Penalty cost for unserved energy (value of lost load) (\$/MWh).}
\nomenclature[B12]{$D^{\text{CO}}_{mc}$}{Material demand of $m$ for producing component $c$ (tonnes/unit).}%
\nomenclature[B13]{$D^{\text{PR}}_{cp}$}{Component demand of $c$ for producing product $p$ (units/MW).}%
\nomenclature[B14]{$\overline{M}_{my}$}{Primary supply limit for material $m$ in year $y$ (tonnes).}%
\nomenclature[B15]{$R^{\text{RM}}_{mg}$}{Recovery rate of material $m$ from retired unit $g$ (tonnes/MW).}%
\nomenclature[B16]{$T^{\text{LEAD}}_g$}{Lead time for deploying unit $g \in \mathcal{G}$ (year).}%
\nomenclature[B17]{$T^{\text{LT}}_g$}{Expected lifetime of unit $g \in \mathcal{G}$ (year).}%
\nomenclature[B18]{$T^{\text{RT}}_g$}{Retirement year for unit $g \in \mathcal{G}$ (year).}%
\nomenclature[B19]{$A^{k}_i$}{Initial available area for deploying technology $k$ in zone $i$ (km\textsuperscript{2}).}%
\nomenclature[B20]{$R^{\text{CAP}}_{k}$}{Capacity density of technology $k$ (MW/km\textsuperscript{2}).}%
\nomenclature[B21]{$C^{\text{I}}_{gy}$}{Capital investment cost for unit $g$ in year $y$ (\$).}%
\nomenclature[B22]{$C^{\text{F}}_{gy}$}{Fix O\&M cost for unit $g$ in year $y$ (\$/MW).}%
\nomenclature[B23]{$C^{\text{V}}_{g}$}{Variable cost of unit $g$ operation (\$/MWh).}%
\nomenclature[B24]{$N_{ty}$}{Number of annual occurrences for day $t$ in year $y$ (days).}%

\nomenclature[C01]{$p_{gthy}$}{Active power generation of unit $g \notin \mathcal{G}^{\text{sto}}$ in day $t$ hour $h$ year $y$ (MW).}%
\nomenclature[C02]{$q_{lthy}$}{Active power flow through corridor $l$ in day $t$ hour $h$ year $y$ (MW).}%
\nomenclature[C03]{$p^{\text{LS}}_{ithy}$}{Load shedding for zone $i$ in day $t$ hour $h$ year $y$ (MW).}%
\nomenclature[C03]{$p^{\text{RM}}_{ithy}$}{Capacity violation for reserve margin for zone $i$ in day $t$ hour $h$ year $y$ (MW).}%
\nomenclature[C04]{$c_{gthy}/dc_{gthy}$}{Charging/discharging power for storage unit $g \in \mathcal{G}^{\text{st}}$ in day $t$ hour $h$ year $y$ (MW).}%
\nomenclature[C05]{$e^{\text{SOC}}_{gthy}$}{State of charge for storage unit $g \in \mathcal{G}^{\text{st}}$ in day $t$ hour $h$ year $y$ (MWh).}%
\nomenclature[C06]{$e^{\text{RPS}}_{ky}$}{Energy violating RPS policy for technology $k$ in year $y$ (MWh).}%
\nomenclature[C07]{$d_{gy}/b_{gy}/r_{gy}/o_{gy}$}{Status of unit $g$ if planned, built, retired, or operational, respectively, in year $y$; binary if $g \in \mathcal{G}^{\text{th}}$, continuous on $[0,1]$ otherwise (unitless).}%
\nomenclature[C11]{$u_{my}$}{Total utilization of material $m$ in year $y$ (metric tonnes).}%
\nomenclature[C12]{$v_{cy}$}{Production of component $c$ in year $y$ (units).}%
\nomenclature[C13]{$w_{py}$}{Production of product $p$ in year $y$ (GW).}%
\nomenclature[C14]{$s_{my}$}{Stock of material $m$ in year $y$ (metric tonnes).}%
\nomenclature[C15]{$f^k_{iy}$}{Available area for technology $k$ in zone $i$ at the beginning of year $y$ (km\textsuperscript{2}).}%

\section{Introduction}
\label{sec:intro}

\subsection{Motivation}

\IEEEPARstart{E}{lectricity} increasingly serves as a critical input to production and as a key indicator of economic development. Extensive empirical evidence highlights a strong link between electricity generation and economic growth. Atems and  Hotaling \cite{atems2018effect} find that electricity generation from various sources supports economic output, with wind and solar technologies showing growing influence in recent years. For example, energy use and Gross Domestic Product (GDP) in the US rose together in the 1990s, while in the 2000s, growth became more aligned with efficient and targeted energy use \cite{arora2016energy}. Recent electrification of transportation and industry and the rise of digital technologies also simultaneously drive economic growth and substantially increase electricity consumption. For instance, full electrification of vehicles in the U.S. could increase electricity demand by about 30\%~\cite{galvin2022electric}, and AI-driven data centers are projected to consume 6.7–12\% of total U.S. electricity by 2028~\cite{shehabi20242024}. These trends underscore the central role of electricity in enabling long-term growth and supporting modern economies. However, the U.S. Department of Energy (DOE) warns that demand is rising faster than  generation and transmission investments, creating new reliability challenges across the national power grid and, in particular, for PJM~\cite{doe2024}. In this context, additional complexity highlighted by \cite{doe2024} is to consider heterogeneous regulatory targets such as state-level renewable portfolio standards (RPS) in multi-state power grids that may affect power grid performance beyond state boundaries. Generation expansion planning (GEP) is  therefore essential for guiding the timing, siting, and scale of generation and infrastructure investments.

\begin{figure*}[!b]
    \centering
    \includegraphics[width=\linewidth]{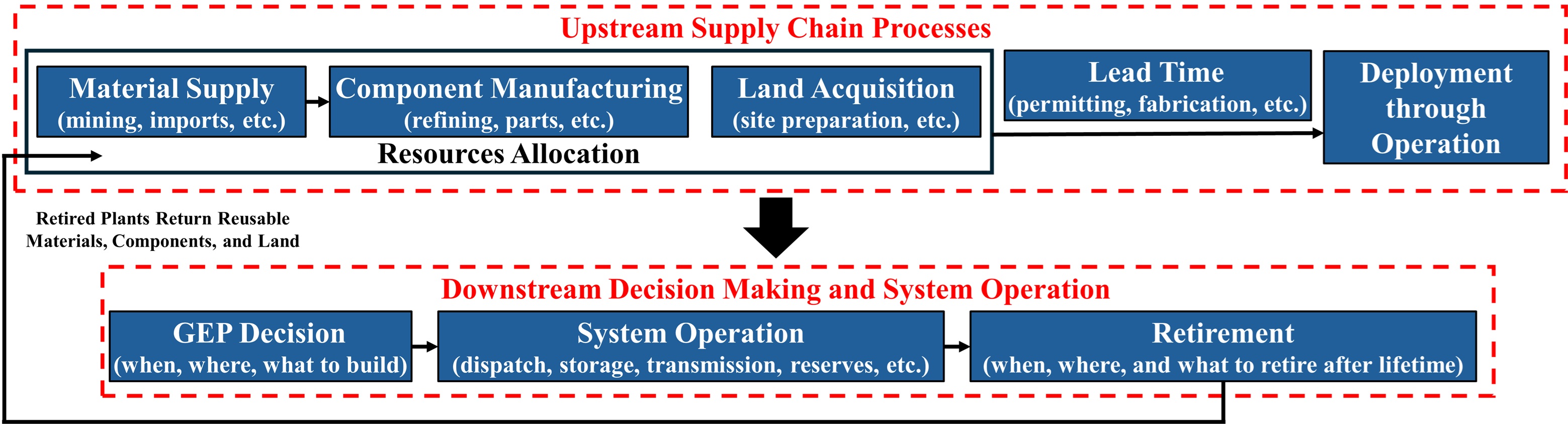}
    \caption{Overview of upstream and downstream components in generation expansion planning}
    \label{fig:diagram}
\end{figure*}

Existing GEP models often focus primarily on downstream decision-making, which typically involve system-level choices (e.g., generation and transmission expansion decisions), which assume that these choices are readily available by the time GEP selects for deployment. In contrast, such upstream components as material supply, manufacturing capabilities, field availability, and permitting (see Fig.~\ref{fig:diagram}) are assumed either unconstrained or reduced to annual capacity limits. Recent studies show that most energy system and capacity expansion models continue to treat material requirements as exogenous parameters, without explicitly linking deployment decisions to upstream supply chain conditions. Schulze et al. \cite{schulze2024overcoming} find that existing modeling frameworks generally rely on externally prescribed material demand trajectories and lack mechanisms to represent material needs endogenously within the optimization process. Similarly, Elshkaki and Shen \cite{elshkaki2019energy} demonstrate that the implications of resource scarcity are frequently underrepresented in long-term energy system modeling, despite their potential to influence the feasibility of low-carbon pathways. As a result, many existing GEP frameworks abstract away or simplify operational and logistical complexities associated with upstream supply chains, potentially overlooking bottlenecks that delay or constrain the deployment of new capacity. These simplifications may undermine the credibility of modeled investment pathways and compromise long-term resource adequacy, especially when infrastructure expansion is time-sensitive or materially intensive.

An increasing number of supply chain disruptions have exposed the risks of overlooking upstream constraints in generation planning. In 2022, U.S. solar installations fell by 17\% year-over-year due to trade barriers and supply bottlenecks, leading to a projected 23\% shortfall in annual deployment~\cite{SEIA2022}. Projects like Ørsted’s Sunrise Wind and the UK’s Morven offshore wind farm have faced major delays due to component shortages and grid connection issues~\cite{WSJ2024, Powell2024MorvenWindFarm}. These are not isolated incidents, but part of a global pattern of geopolitical and logistical uncertainty threatening project timelines. As modern energy systems rely on materials- and component-intensive technologies, integrating upstream supply chain constraints into generation expansion planning is essential for anticipating deployment risks, mitigating delays, and supporting realistic long-term capacity decisions.

\subsection{Literature Review}
\label{Literaure}

Understanding why upstream supply chain constraints have been largely absent from traditional GEP models provides important context for this study. The intent is not to examine or critique traditional GEP models, but to clarify the modeling gap that motivates the need for integrating supply chain constraints.


First, the long-standing stability and affordability of supply chains for traditional energy infrastructure assets led to the perception that upstream constraints were not a serious concern. For decades, coal, natural gas, and nuclear plants were built using established industrial materials like steel and cement that were widely available, low in cost, and typically made up less than five percent of total capital expenditures, even during commodity price spikes~\cite{rong2012does}. Favorable financial performance and relatively predictable construction processes further used to reinforce the assumption that supply chain issues would not materially affect project viability~\cite{iea2024coal}. As a result, early GEP frameworks were designed around the dominant risks of the past times, including fuel prices, reliability, and emissions, rather than upstream manufacturing or material limitations.

Nevertheless, the supply chain requirements of modern energy technologies differ markedly from those of traditional assets. Technologies such as wind turbines, photovoltaic systems, and battery storage depend on globally distributed supply chains and specialized components manufactured off-site. Their deployment is therefore sensitive to material availability, manufacturing throughput, transportation logistics, and coordination across multiple supplier tiers. Recent disruptions, including those observed during the COVID-19 pandemic, exposed vulnerabilities in these increasingly complex and internationally interconnected supply networks~\cite{john2024supply}. These developments have shifted upstream supply chain considerations from a background assumption to a central factor influencing deployment timelines.

Second, although some modeling frameworks address elements of the supply chain, integration of upstream constraints into GEP is limited. Addressing these dimensions requires interdisciplinary methods bridging energy systems, industrial engineering, and trade. Existing tools for supply chain and resource availability analysis are often siloed or lack the temporal, spatial, and operational granularity needed for power system planning. For example, Integrated Assessment Models (IAMs) include long-term resource constraints and project energy transitions at global or national scales, but their coarse spatial resolution and multi-year steps preclude detailed assessments of short-term system performance or localized bottlenecks. In contrast, Life Cycle Assessment (LCA) tools offer detailed insights into the environmental and material footprints of technologies but are typically retrospective, static, and not suited for prospective system decision-making. Few frameworks offer a unified approach to embed upstream supply chain dynamics into operationally relevant generation planning. Table~\ref{tab:model_comparison} summarizes key features and limitations of IAMs, LCAs, and traditional GEP models.

\begin{table}[t]
\caption{Comparison of IAM, LCA, and Traditional GEP.}
\label{tab:model_comparison}
\centering
\renewcommand{\arraystretch}{1.2}
\large
\resizebox{\linewidth}{!}{
\LARGE
\begin{tabular}{|>{\centering\arraybackslash}p{3.7cm} 
                |>{\centering\arraybackslash}p{4cm} 
                |>{\centering\arraybackslash}p{3.8cm} 
                |>{\centering\arraybackslash}p{4.3cm}|}
\hline
& IAM~\cite{qiu2024impacts, wilkerson2015comparison} & LCA~\cite{pehnt2006dynamic, zhang2023graphite} & Trad. GEP~\cite{lara2018deterministic}\\
\hline
Objective & Global systems view & Material intensity & Optimal planning \\
\hline
Temporal & Long-term, coarse & Static or lifetime & Hourly to annual \\
\hline
Decision Logic & Recursive-logit shares & No decision logic & Inter-temporal optimization \\
\hline
Supply Chain & Simple resource curves & Per-unit only & Mostly ignored \\
\hline
System Scope & Cross-sectoral & Tech-level only & Power systems \\
\hline
Limitations & Low resolution, stylized & Narrow scope, static & No upstream limits \\
\hline
\end{tabular}
}
\end{table}

Third, upstream supply chain constraints remain complex to formal modeling due to their inherent uncertainty and the lack of standardized data. While downstream elements such as load, generation, transmission, and pricing are well-documented and supported by established modeling tools, upstream factors like material availability, production capacity, and lead times are more difficult to quantify and verify. This gap has made it challenging to develop appropriate formal modeling and thus to  incorporate upstream dynamics into GEP.

Although recent work has begun to highlight the importance of upstream constraints. Zhang et al.\cite{zhang2022global} quantify global supply risks for metals critical to clean energy, revealing vulnerabilities in material availability, and \cite{patankar2023land} analyze land-use trade-offs in electricity decarbonization across the American West, underscoring spatial limitations in infrastructure siting. Yet these analysis are not integrated into GEP frameworks, which is needed to synchronize infrastructure delivery with future demand and to ensure that modeled expansion pathways remain feasible under real-world supply chain conditions.

\subsection{Contributions}

This paper makes the following contributions to address the gap identified in Section~\ref{Literaure}:

\begin{itemize}

    \item We develop a Supply-Chain-Constrained Generation Expansion Planning (SC-GEP) model that explicitly incorporates material procurement, manufacturing capacity, spatial deployment limitations, and technology-specific lead times into a unified multi-stage GEP framework.

    \item We demonstrate that upstream supply-chain frictions increasingly determine the pace and feasibility of clean-energy deployment, and we evaluate how these constraints influence investment timing, system reliability, and operational flexibility.

    \item We ground the modeling need in national resource adequacy concerns regarding energy demand-deployment mismatches~\cite{doe2024}. Maryland serves as the case study, characterized by high PJM reliability risks and the Abundant Affordable Clean Energy (AACE)  Act~\cite{md_sb316_2025}, which mandates substantial renewable growth while restricting new gas development.

    \item We show that the SC-GEP model can identify material bottlenecks, anticipate deployment delays, and support more resilient and reliable capacity expansion decisions under upstream supply chain constraints.

\end{itemize}

\section{Mathematical formulation}
\label{sec:formulation}

This section presents the mathematical formulation of the proposed SC-GEP model, which optimizes generation expansion decisions while accounting for upstream supply chain constraints, including material availability, component and product assembly, deployment lead times, and spatial requirements. The model ensures feasible lead times for deployment of generation resources by capturing critical interactions between supply chain limitations and infrastructure planning.

The SC-GEP model comprises two integrated components: (1) a supply chain (SC) module and (2) a generation expansion planning (GEP) module, which together identify when and where to invest in new infrastructure to meet projected electricity demand. Rather than arguing for a specific supply chain or GEP model, this work aims to emphasize the importance of explicitly linking supply chain constraints with generation expansion planning. Mathematically, the model is formulated as a multi-stage mixed-integer linear program (MILP) involving discrete investment and retirement decisions and continuous operational variables across multiple time periods. To improve computational efficiency, the model is solved using Nested Benders Decomposition (NBD), which partitions it into stage-wise subproblems coordinated by iteratively updated Benders cuts, as detailed in Section~\ref{sec:algorithm}. Unless stated otherwise, summations and $\forall$ statements span the full set of indexed elements.

\subsection{Objective Function}

The objective function in~\eqref{eq:objective} includes three cost components. First, investment cost \( C^{\text{in}}_y \) covers capacity expansion for new generators and is computed using the adjusted and discounted cost \( AC^{\text{I}}_{gy} \)\footnote{\( AC^{\text{I}}_{gy} \) denotes the present value of capital investment for unit \( g \) in year \( y \), adjusted to reflect only the effective years of operation within the planning horizon. If a unit’s construction is delayed due to lead time or its lifetime extends beyond the final year of the model, the investment cost is prorated accordingly and discounted to its net present value.}. Second, operational cost \( C^{\text{op}}_y \) includes fixed operation and maintenance costs \( C^{\text{F}}_g \) based on installed capacity, variable generation costs \( C^{\text{V}}_g \) weighted by hourly dispatch for thermal and renewable units, and storage costs for both charging and discharging, evaluated using \( C^{\text{V}}_g \) per MWh. Third, penalty cost \( C^{\text{pe}}_y \) is the costs of unserved load valued at the Value of Lost Load (VOLL) \( P^{\text{VOLL}} \), reserve margin and RPS shortfalls penalized at \( P^{\text{RM}} \) and \( P^{\text{RPS}} \).

\begin{subequations}
\label{eq:objective_group}

\begin{equation}
\min \sum_{y} C_y = \sum_{y} \left( C^{\text{in}}_y + C^{\text{op}}_y + C^{\text{pe}}_y \right)
\label{eq:objective}
\end{equation}

\begin{equation}
C^{\text{in}}_y = \sum_{g \in \tilde{\mathcal{G}}} AC^{\text{I}}_{gy} \cdot \overline{P}^{\text{G}}_g \cdot d_{gy}
\label{eq:investment_cost}
\end{equation}

\begin{equation}
\begin{split}
C^{\text{op}}_y = \sum_{g} C^{\text{F}}_g \cdot \overline{P}^{\text{G}}_g \cdot o_{gy}
+ \sum_{g \in \mathcal{G}^{\text{th}} \cup \mathcal{G}^{\text{rn}}} \sum_{t} C^{\text{V}}_g \cdot N_{ty} \cdot \sum_{h} p_{gthy} \\
+ \sum_{g \in \mathcal{G}^{\text{st}}} \sum_{t} C^{\text{V}}_g \cdot N_{ty} \cdot \sum_{h} \left( c_{gthy} + dc_{gthy} \right)
\end{split}
\label{eq:operation_cost}
\end{equation}

\begin{equation}
C^{\text{pe}}_y = \sum_{t} N_{ty} \cdot P^{\text{VOLL}} \cdot \sum_{h} p^{\text{LS}}_{ithy} 
+ P^{\text{RM}} \cdot p^{\text{RM}}_{y} 
+ P^{\text{RPS}} \cdot e^{\text{RPS}}_{ky}
\label{eq:penalty_cost}
\end{equation}

\end{subequations}

\subsection{Supply Chain (SC) Module}

A conventional GEP formulation does not model upstream supply chains and implicitly assumes that capacity can be built whenever it is economically optimal. This leads to the treatment of supply as perfectly elastic and disregars material availability, lead times, and spatial deployment limits. As a result, the traditional GEP model cannot factor in the effect of deployment delays, shortages, or infeasible build schedules on its decisions.

On the other hand, the SC module introduced in this study explicitly models upstream constraints, including materials, components, products, lead times, and field availability, to capture their direct impact on the feasibility and timing of capacity deployment. These constraints can delay or restrict new builds, trigger reserve margin violations, or increase unserved load penalties, thereby altering both investment decisions and operational outcomes.

\begin{subequations}
\label{eq:sc_constraints_all}

\begin{equation}
u_{my} \geq \sum_{c} v_{cy} \cdot D^{\text{CO}}_{mc}, 
\quad \forall m, y
\label{eq:raw_material_demand}
\end{equation}

\begin{equation}
v_{cy} \geq \sum_{p} w_{py} \cdot D^{\text{PR}}_{cp}, 
\quad \forall c, y
\label{eq:component_production}
\end{equation}

\begin{equation}
u_{my} \leq \overline{M}_{my} 
+ \sum_{g \in \mathcal{G}} R^{\text{RM}}_{mg} \cdot \overline{P}^{\text{G}}_g \cdot r_{gy}
+ s_{my}, 
\quad \forall m, y
\label{eq:material_supply}
\end{equation}

\begin{equation}
\begin{split}
s_{my} = s_{m(y-1)} + \overline{M}_{m(y-1)} - u_{m(y-1)} \\
+ \sum_{g \in \mathcal{G}} R^{\text{RM}}_{mg} \cdot \overline{P}^{\text{G}}_g \cdot r_{g(y-1)}, 
\quad \forall m, y
\label{eq:stock_balance}
\end{split}
\end{equation}

\begin{equation}
\sum_{g \in \mathcal{G}^{k}} \overline{P}^{\text{G}}_g \cdot d_{gy} 
\leq \sum_{p \in \mathcal{P}^{k}} w_{py}, 
\quad \forall k, y
\label{eq:product_capacity}
\end{equation}

\begin{equation}
\frac{\sum_{g \in \mathcal{G}^{k} \cap \mathcal{G}_{i}} 
\overline{P}^{\text{G}}_g \cdot d_{gy}}{R^{\text{CAP}}_{k}} 
\leq f^k_{iy}, 
\quad \forall i, y, k
\label{eq:field_use}
\end{equation}

\begin{equation}
\begin{split}
f^k_{iy} = f^k_{i(y-1)} 
+ \frac{\sum_{g \in \mathcal{G}^{k} \cap \mathcal{G}_{i}} 
\overline{P}^{\text{G}}_g \cdot r_{gy}}{R^{\text{CAP}}_{k}} \\
- \frac{\sum_{g \in \mathcal{G}^{k} \cap \mathcal{G}_{i}} 
\overline{P}^{\text{G}}_g \cdot d_{g(y-1)}}{R^{\text{CAP}}_{k}}, 
\quad \forall i, y > 1, k
\label{eq:field_status_all}
\end{split}
\end{equation}

\begin{equation}
f^k_{i1} = A^k_i 
+ \frac{\sum_{g \in \mathcal{G}^{k} \cap \mathcal{G}_{i}} 
\overline{P}^{\text{G}}_g \cdot r_{g1}}{R^{\text{CAP}}_{k}}, 
\quad \forall i, k
\label{eq:initial_field}
\end{equation}

\begin{equation}
\sum_{y'' \leq y - T^{\text{LEAD}}_g} d_{g y''} 
= \sum_{y' \leq y} b_{g y'}, 
\quad \forall g \in \tilde{\mathcal{G}},\; y
\label{eq:lead_time}
\end{equation}

\begin{equation}
b_{gy} = 0, 
\quad \forall g \in \bar{\mathcal{G}},\; y
\label{eq:existing_no_built}
\end{equation}

\begin{equation}
\sum_{y'' \leq y - T^{\text{LT}}_g} b_{g y''} 
= \sum_{y' \leq y} r_{g y'}, 
\quad \forall g \in \tilde{\mathcal{G}},\; y
\label{eq:lifetime_retirement}
\end{equation}

\begin{equation}
r_{gy} = 1 \; \text{if } y = T^{\text{RT}}_g,\; 0 \; \text{if } y < T^{\text{RT}}_g, 
\quad \forall g \in \bar{\mathcal{G}},\; \forall y
\label{eq:existing_retirement}
\end{equation}

\end{subequations}

Eqs.~\eqref{eq:raw_material_demand}--\eqref{eq:component_production} ensure that material and component availability meets downstream production needs. Eq.~\eqref{eq:raw_material_demand} ensures sufficient material inputs for component manufacturing, while Eq.~\eqref{eq:component_production} ensures adequate component production to support product assembly.
Eqs.~\eqref{eq:material_supply}--\eqref{eq:stock_balance} govern material acquisition and stock dynamics. Eq.~\eqref{eq:material_supply} requires that material demand be met through a mix of primary supply, recovered materials from retired units, and stock. Eq.~\eqref{eq:stock_balance} tracks annual stock levels based on inflows and outflows. Eq.~\eqref{eq:product_capacity} limits total capacity expansion by the availability of manufactured products. Eqs.~\eqref{eq:field_use}--\eqref{eq:field_status_all} manage field resource use and replenishment. Eq.~\eqref{eq:field_use} restricts annual field use based on available area and capacity density, while Eq.~\eqref{eq:field_status_all} updates the available area annually, accounting for previous use and field returns. Eq.~\eqref{eq:initial_field} defines initial area availability at the start of the planning horizon. Eqs.~\eqref{eq:lead_time}--\eqref{eq:existing_no_built} enforce lead-time constraints, allowing new units to operate only after \( T^{\text{LEAD}}_g \) years and excluding existing units from decisions to build. Eqs.~\eqref{eq:lifetime_retirement}--\eqref{eq:existing_retirement} impose retirement conditions: candidate units retire after their design lifetime \( T^{\text{LT}}_g \), while existing units retire at a fixed year \( T^{\text{RT}}_g \), with no early or delayed retirements allowed. Together, these constraints capture the full supply chain dynamics from resource allocation and lead-time-driven deployment to eventual retirement.

\subsection{Generation Expansion Planning (GEP) Module}

The GEP module includes two sets of constraints. The first set (\eqref{eq:energy_balance}–\eqref{eq:tech_specific_rps}) covers system operations including power balance, generation limits, transmission flows, load shedding, reserve margins, and RPS compliance. The second set (\eqref{eq:storage_charging}–\eqref{eq:daily_balance}) covers storage operations, including charging/discharging limits, state-of-charge (SOC) dynamics, and daily energy balancing. This separation clarifies the distinct operational role of storage as energy-limited resources.

\subsubsection{System Operation Constraints}

\begin{subequations}
\label{eq:gep_constraints_all}

\begin{equation}
\begin{split}
\sum_{g \in \mathcal{G}_{i} \cap (\mathcal{G}^{\text{th}} \cup \mathcal{G}^{\text{rn}})} p_{gthy} 
+ \sum_{g \in \mathcal{G}_{i} \cap \mathcal{G}^{\text{st}}} \left(dc_{gthy} - c_{gthy}\right) 
- \sum_{l \in \mathcal{LS}_{i}} q_{lthy} \\
+ \sum_{l \in \mathcal{LR}_{i}} q_{lthy} = L_{ithy} - p_{ithy}^{\text{LS}}, 
\quad \forall i, t, h, y
\label{eq:energy_balance}
\end{split}
\end{equation}

\begin{equation}
0 \leq p_{gthy} \leq \overline{P}^{\text{G}}_g \cdot o_{gy}, 
\quad \forall g \in \mathcal{G}^{\text{th}},\; y,t,h
\label{eq:gen_output_thermal}
\end{equation}

\begin{equation}
0 \leq p_{gthy} \leq F^{\text{GEN}}_{igthy} \cdot \overline{P}^{\text{G}}_g \cdot o_{gy}, 
\quad \forall g \in \mathcal{G}_i \cap \mathcal{G}^{\text{rn}},\; i,y,t,h
\label{eq:gen_output_renewable}
\end{equation}

\begin{equation}
o_{gy} = o_{g(y-1)} + b_{gy} - r_{gy}, 
\quad \forall g \in \mathcal{G},\; y \geq 1
\label{eq:operational_status}
\end{equation}

\begin{equation}
o_{g0} = 1\; \text{if } g \in \bar{\mathcal{G}},\; 0\; \text{otherwise}, 
\quad \forall g \in \mathcal{G}
\label{eq:initial_condition}
\end{equation}

\begin{equation}
-\overline{P}^{\text{L}}_l \leq q_{lthy} \leq \overline{P}^{\text{L}}_l, 
\quad \forall l, y, t, h
\label{eq:transmission}
\end{equation}

\begin{equation}
\sum_{g \in \mathcal{G}^{k}} \overline{P}^{\text{G}}_g \cdot F^{\text{ELCC}}_{ky} \cdot o_{gy}
+ p^{\text{RM}}_y \geq (1 + R^{\text{RM}}_y) \cdot \overline{L}_y, 
\quad \forall y, k
\label{eq:reserve_margin}
\end{equation}

\begin{equation}
0 \leq p_{ithy}^{\text{LS}} \leq L_{ithy}, 
\quad \forall i, y, t, h
\label{eq:load_shedding}
\end{equation}

\begin{equation}
\begin{split}
\sum_{g \in \mathcal{G}^{k}} \sum_{t} \sum_{h} 
N_{ty} \cdot p_{gthy}  
+ e^{\text{RPS}}_{ky} \\
\geq R^{\text{RPS}}_{ky} \cdot \sum_{t} \sum_{h} \sum_{i} 
N_{ty} \cdot L_{ithy}, 
\quad \forall k, y
\end{split}
\label{eq:tech_specific_rps}
\end{equation}

\end{subequations}

Eq.~\eqref{eq:energy_balance} enforces nodal power balance by ensuring that total generation, net imports from adjacent zones, and storage satisfy the local demand minus load shedding. Eqs.~\eqref{eq:gen_output_thermal}--\eqref{eq:gen_output_renewable} limit thermal and renewable generation based on installed capacity, operational status, and renewable availability factors. Eqs.~\eqref{eq:operational_status}--\eqref{eq:initial_condition} define the year-to-year evolution of operational status based on prior-year status and current build or retirement decisions, with existing units (\( \bar{\mathcal{G}} \)) assumed operational at the start and candidate units (\( \tilde{\mathcal{G}} \)) initially inactive. Eq.~\eqref{eq:transmission} constrains power flow within rated line capacity in both directions. Eq.~\eqref{eq:reserve_margin} maintains system reliability by requiring effective load-carrying capability (ELCC)-adjusted available capacity to meet peak demand plus the planning reserve margin, with slack variable $p^{\text{RM}}_{y}$ for shortfalls. Eq.~\eqref{eq:load_shedding} restricts load shedding to be non-negative and not exceed a given demand. Finally, Eq.~\eqref{eq:tech_specific_rps} enforces technology-specific RPS compliance by requiring annual generation from each technology to meet the mandated share of system demand, with slack variable $e^{\text{RPS}}_{ky}$ capturing any shortfall.

\subsubsection{Storage Constraints}

\begin{subequations}
\label{eq:storage_constraints_all}

\begin{equation}
0 \leq c_{gthy} \leq \overline{P}^{\text{G}}_g \cdot o_{gy}, 
\quad \forall g \in \mathcal{G}^{\text{st}},\; y,t,h 
\label{eq:storage_charging}
\end{equation}

\begin{equation}
0 \leq dc_{gthy} \leq \overline{P}^{\text{G}}_g \cdot o_{gy}, 
\quad \forall g \in \mathcal{G}^{\text{st}},\; y,t,h 
\label{eq:storage_discharging}
\end{equation}

\begin{equation}
0 \leq e^{\text{soc}}_{gthy} \leq \overline{E}^{\text{ST}}_g, 
\quad \forall g \in \mathcal{G}^{\text{st}},\; y,t,h
\label{eq:soc_limit}
\end{equation}

\begin{equation}
e^{\text{soc}}_{gthy} = e^{\text{soc}}_{gt(h-1)y} 
+ \epsilon^{\text{CH}} \cdot c_{gthy} 
- \frac{dc_{gthy}}{\epsilon^{\text{DC}}}, 
\quad \forall g \in \mathcal{G}^{\text{st}},\; y,t,h 
\label{eq:storage_dynamics}
\end{equation}

\begin{equation}
e^{\text{soc}}_{gt1y} = e^{\text{soc}}_{gt\text{end}y} 
= 0.5 \cdot \overline{E}^{\text{ST}}_g \cdot o_{gy}, 
\quad \forall g \in \mathcal{G}^{\text{st}},\; y,t
\label{eq:daily_balance}
\end{equation}

\end{subequations}

Eqs.~\eqref{eq:storage_charging}--\eqref{eq:storage_discharging} limit charging and discharging power to rated capacity when operational. Eq.~\eqref{eq:soc_limit} bounds the SOC within the installed energy capacity, while Eq.~\eqref{eq:storage_dynamics} updates SOC based on previous levels, charging (adjusted for efficiency), and discharging. Eq.~\eqref{eq:daily_balance} ensures energy neutrality by requiring the SOC to return to 50\% of capacity at the start and end of each representative day~\cite{7902215}.

\section{Decomposition for Accelerating Solving}
\label{sec:algorithm}

The multi-stage MILP in Section~\ref{sec:formulation} can be solved with commercial solvers but becomes computationally demanding with high temporal resolution and detailed supply chain layers. To address this, we adopt a nested Benders decomposition framework, following~\cite{zou2019stochastic} and its extension in~\cite{lara2018deterministic} to accommodate both continuous and binary state variables. Although only Lagrangian cuts guarantee finite convergence and eliminate duality gaps (at the expense of a more complex reformulation and subgradient optimization), we implement standard Benders cuts due to their simplicity, computational efficiency, and strong alignment with the supply chain-driven structure of SC-GEP, where capturing supply chain dynamics is prioritized over algorithmic tightness. This in tern enables to solve each iteration of the decomposition faster, than with Lagrangian cuts, but may require a larger number of iterations. 

For clarity of exposition, we express the original multi-stage problem in the following compact form:

\begin{subequations}
\label{eq:expanded_original_problem}
\begin{align}
\min_{\{m_y, n_y\}_{y \in \mathcal{Y}}} 
&\quad \sum_{y \in \mathcal{Y}} f_y(m_y, n_y) \label{eq:expanded_original_problem_obj} \\
\text{s.t.} \quad 
& A_y m_y + B_y n_y \leq b_y, && \forall y \in \mathcal{Y} 
\label{eq:expanded_original_problem_operational} \\
& C_1 m_1 \leq f_1, 
\label{eq:expanded_original_problem_initial} \\
& C_{y-1} m_{y-1} + D_y m_y \leq f_y, && \forall y = 2, \ldots, |\mathcal{Y}| 
\label{eq:expanded_original_problem_coupling} \\
& m_y \in \mathcal{M}_y, \quad n_y \in \mathcal{N}_y, && \forall y \in \mathcal{Y}
\label{eq:expanded_original_problem_feasibility}
\end{align}
\end{subequations}

Here, \( m_y \) and \( n_y \) denote cross-stage (\( s_{my}, f^k_{iy}, d_{gy}, b_{gy}, o_{gy} \)) and stage-wise variables (e.g., other variables for supply chain, operations, and storage), respectively. The objective~\eqref{eq:expanded_original_problem_obj} corresponds to Eq.~\eqref{eq:objective_group}, minimizing total system cost across all years. Constraints~\eqref{eq:expanded_original_problem_operational} represent stage-wise operational limits, including Eqs.~\eqref{eq:raw_material_demand}–\eqref{eq:material_supply}, \eqref{eq:product_capacity}, \eqref{eq:field_use}, \eqref{eq:energy_balance}–\eqref{eq:gen_output_renewable}, and \eqref{eq:transmission}–\eqref{eq:tech_specific_rps}. The initial values of cross-stage variables (e.g., \( s_{my} \), \( f^k_{iy} \), \( o_{gy} \), \( b_{gy} \), \( d_{gy} \)) are set via~\eqref{eq:expanded_original_problem_initial}, including Eqs.~\eqref{eq:initial_field}, \eqref{eq:existing_no_built}, \eqref{eq:existing_retirement}, and \eqref{eq:initial_condition}. Cross-stage consistency is enforced by~\eqref{eq:expanded_original_problem_coupling}, which includes Eqs.~\eqref{eq:stock_balance}, \eqref{eq:field_status_all}, \eqref{eq:lead_time}, \eqref{eq:lifetime_retirement}, and \eqref{eq:operational_status}. Finally, feasibility is ensured by~\eqref{eq:expanded_original_problem_feasibility}.

Decomposition is enabled by reformulating the cross-stage constraints~\eqref{eq:expanded_original_problem_coupling} using continuous duplicated variables \( z_y \) to isolate each year’s subproblem. The reformulated constraint is compactly expressed as:

\begin{subequations}
\label{eq:cross_stage_reformulation}
\begin{align}
C_{y-1} z_y + D_y m_y &\leq f_y, 
&& \forall y = 2, \ldots, |\mathcal{Y}| 
\label{eq:cross_stage_subproblem} \\[0.5ex]
(\mu_y): z_y &= \widehat{m}_{y-1}, 
&& \forall y = 2, \ldots, |\mathcal{Y}|
\label{eq:cross_stage_linking}
\end{align}
\end{subequations}

Here, \( z_y \) are duplicated state variables linked to the forward-pass solution \( \widehat{m}_{y-1} \) from the previous stage. This reformulation enables stage-wise decomposition. The resulting subproblem for year~\( y \) at iteration~\( \nu \) is:

\begin{equation}
\begin{aligned}
    C_{y\nu} (\hat{m}_{(y-1)\nu}, \phi_{y\nu}) 
    &= \min_{m_y, n_y} \; f_y(m_y, n_y) + \phi_{y\nu}(m_y) \\
    \text{s.t.} \quad 
    & A_y m_y + B_y n_y \leq b_y, \\
    & C_{y-1} z_y + D_y m_y \leq f_y,\\
    & (\mu_y): \quad z_y = \hat{m}_{(y-1)\nu}, \\
    & m_y \in \mathcal{M}_y, \quad n_y \in \mathcal{N}_y
\end{aligned}
\label{eq:yearly_problem_benders}
\end{equation}

The cost-to-go function \( \phi_{y\nu}(m_y) \) is approximated using accumulated Benders cuts:

\begin{equation}
\phi_{y\nu}(m_y)
:= 
\min_{m_y,\, \alpha_y}
\left\{
\alpha_y :
\begin{aligned}
\alpha_y &\ge 
\hat{C}_{(y+1)\nu} 
\\
&\quad +\, \mu_{(y+1)\nu}^\top (\hat{m}_{y\nu} - m_y)
\end{aligned}
\right\}
\label{eq:cost_to_go}
\end{equation}

Here, \( \alpha_y \) represents the approximated cost-to-go for stage~\( y \), and \( \mu_{(y+1)\nu} \) are duals from the linking constraint~\eqref{eq:cross_stage_linking}, capturing the sensitivity of future costs to cross-stage variables. Algorithm~\ref{alg:decomposition} outlines the full procedure.

\setlength{\textfloatsep}{6pt}
\begin{algorithm}[h]
\caption{Decomposition Algorithm for SC-GEP.}
\label{alg:decomposition}
\begin{algorithmic}[1]
\STATE \textbf{Init:} $\hat{m}_{0,0}$, $\mathcal{B}_y \gets \emptyset$, $LB_0 \gets -\infty$, $UB_0 \gets +\infty$, $\nu \gets 0$
\WHILE{$UB_\nu - LB_\nu > \epsilon$ and $\nu < \nu_{\max}$}
    \FOR{\textbf{(Forward Pass)} $y = 1,\dots,Y$}
        \STATE Solve $C_{y\nu} = \min f_y(m_y, n_y) + \phi_{y\nu}(m_y)$ s.t. $z_y = \hat{m}_{(y-1)\nu}$; Store $\hat{m}_{y\nu},\; \hat{C}_{y\nu}$
    \ENDFOR
    \STATE $UB_\nu \gets \sum_y \hat{C}_{y\nu}$
    \FOR{\textbf{(Backward Pass)} $y = Y,\dots,1$}
        \STATE Solve relaxed LP at stage $y$ for dual $\mu_{y\nu}$; add cut: $\phi_{(y-1)\nu}(m_{y-1}) \geq \hat{C}_{y\nu} + \mu_{y\nu}^T (m_{y-1} - \hat{m}_{(y-1)\nu})$ to $\mathcal{B}_{y-1}$
    \ENDFOR
    \STATE Solve $\min f_1(x_1) + \phi_{1\nu}(x_1)$ over $\mathcal{B}_1$; $LB_\nu \gets$ objective value; $\nu \gets \nu + 1$
\ENDWHILE
\STATE \textbf{Output:} $\{m^*_y, n^*_y, z^*_y\},\; \sum_y C^*_y$
\end{algorithmic}
\end{algorithm}

\section{Case Study}
\label{sec:casestudy}

This case study applies the proposed SC-GEP model to Maryland, a net importer of electricity within the PJM that relies on neighboring PJM states to meet its demand. The state’s AACE Act (HB0398/SB0316) \cite{md_sb316_2025} outlines support for electricity sector investments, including renewable generation and energy storage, while placing restrictions on new natural gas infrastructure. As Maryland continues to promote generation expansion through modern technologies tied to globally constrained supply chains, it provides a timely context to assess upstream limitations. This study focuses on two key questions: (i) how supply chain constraints shape optimal investment planning, and (ii) how overlooking these constraints can result in unrealistic pathways, particularly under growing demand from data centers and electrification, threatening reliability and implementation success. While the SC-GEP model is designed to support decision-making by policymakers and stakeholders, this case study is exploratory in nature and does not seek to resolve specific policy debates. Such applications can be pursued in future work using the proposed methods.


The Maryland power system is represented using four utility service zones—BGE, APS, DPL, and PEPCO—each modeled as a single node, consistent with the zonal framework in~\cite{dvorkinenergy}. Within each zone, renewable generators are aggregated and assigned a common availability profile. Investment decisions are evaluated on an annual basis, while system operations are modeled at an hourly resolution using one representative day per season to capture seasonal and diurnal variations in load and renewable output. These representative days are selected through $k$-means clustering~\cite{fazlollahi2014multi} and retain full 24-hour profiles. Seasons are defined as Spring (Mar–Jun), Summer (Jun–Sep), Fall (Sep–Dec), and Winter (Dec–Mar).

\begin{figure}[t]
    \centering
    \includegraphics[width=1\linewidth]{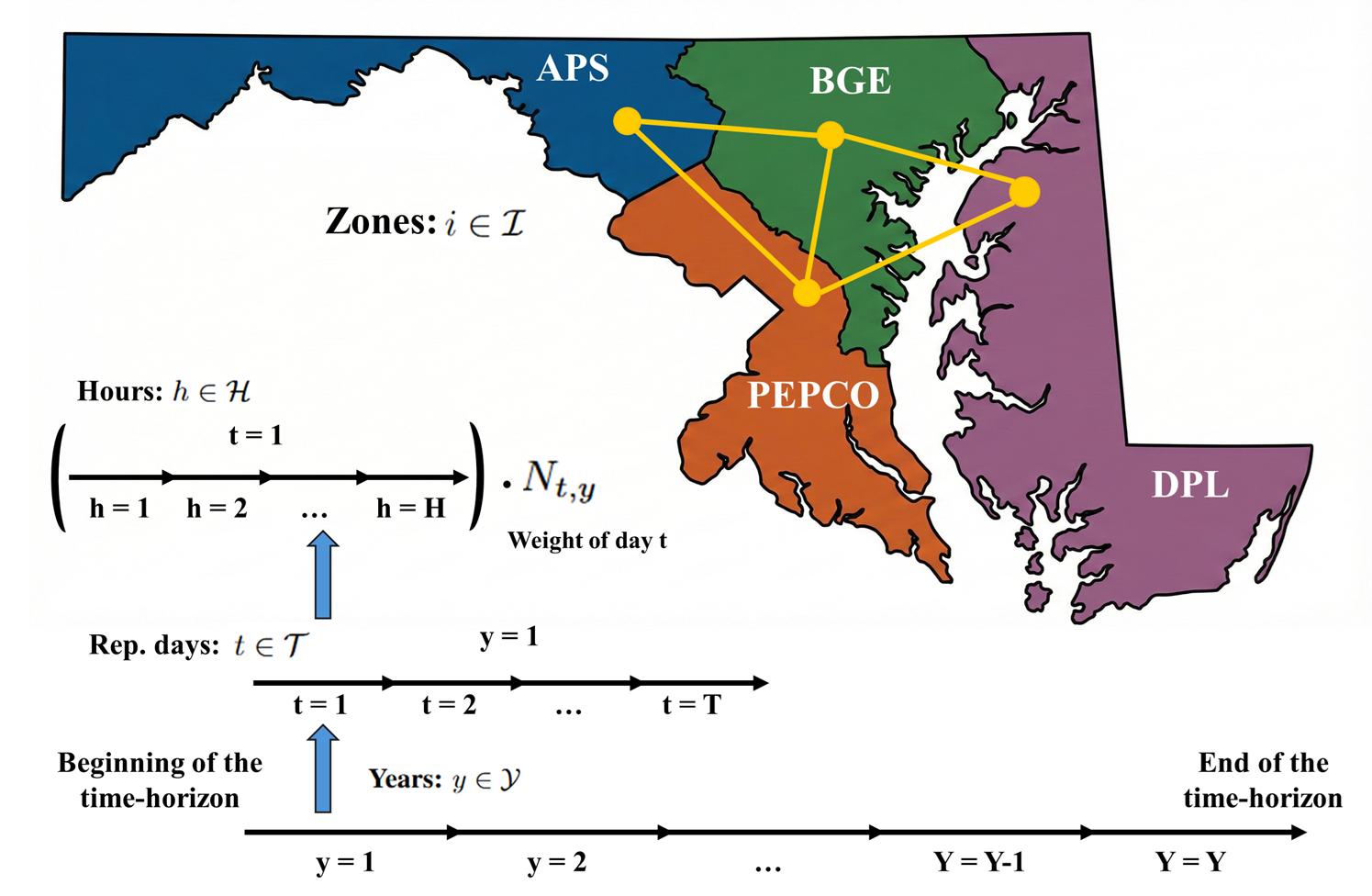}
    \caption{Spatial and temporal representation for maryland power system.}
    \label{fig:spacial_temporal}
\end{figure}

The transmission network is simplified to a four-node model representing Maryland’s service zones. PJM Window 3 upgrades~\cite{pjm2022rtep} are assumed to be fully online, with no additional expansion considered. Power transfers follow a transportation model~\cite{short2011regional} that ignores reactive power and assumes negligible losses. While this abstraction omits effects such as congestion and voltage support~\cite{munoz2013approximations}, it does not detract from the study’s focus on supply chain impacts in long-term planning. Figure~\ref{fig:spacial_temporal} illustrates the spatial (top) and temporal (bottom) structure of the case study, including a simplified single-line diagram of the four-zone network shown in yellow on the map.

\subsection{Supply Chain Representation}
\label{sec:supplychain}

The SC-GEP model captures three supply chain dimensions: material flow, lead times, and field availability.

\subsubsection{Material Flow}

The material flow includes material acquisition, component manufacturing, and final product assembly, with many components shared across products. We model 14 critical materials identified by United States Geological Survey (USGS) and DOE \cite{applegate2023final, igogo2022america}: aluminum, cobalt, dysprosium, gallium, graphite, lithium, manganese, neodymium, nickel, praseodymium, silicon, terbium, tin, and titanium. Material-to-component and component-to-product mappings, along with material demand data, follow previously published work \cite{dvorkin2024understanding}, which details 11   key products used in wind, solar, and battery systems. While not all life-cycle materials are modeled, the selected ones are assumed to be critical, and others are treated as sufficiently available.

\begin{figure}[t]
    \centering
    \includegraphics[width=1\linewidth]{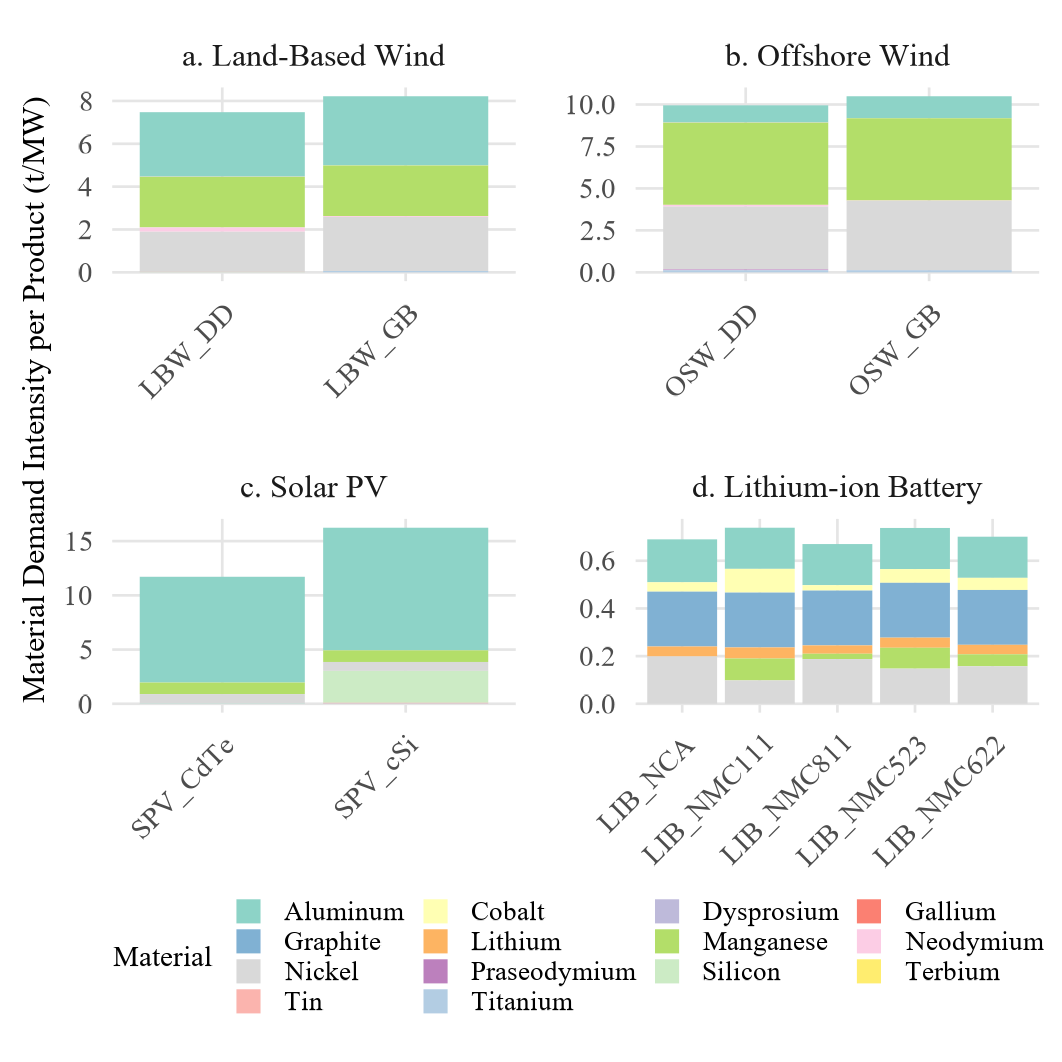}
    \caption{Material intensity (tonnes of materials per MW of installed capacity) for selected products across technologies.}
    \label{fig:material}
\end{figure}

Figure~\ref{fig:material} illustrates the material demand intensity (t/MW) for selected products across land-based wind, offshore wind, and solar PV, as well as representative lithium-ion battery chemistries. 
Abbreviations used in the figure include land-based wind with direct-drive generators (\textnormal{LBW\_DD}) and land-based wind with gearbox generators (\textnormal{LBW\_GB}); offshore wind with direct-drive generators (\textnormal{OSW\_DD}) and offshore wind with gearbox generators (\textnormal{OSW\_GB}); solar PV utility-scale cadmium-telluride systems (\textnormal{SPV\_CdTe}) and solar PV crystalline-silicon systems (\textnormal{SPV\_cSi}); lithium-ion batteries using nickel-cobalt-aluminum cathodes (\textnormal{LIB\_NCA}), and nickel-manganese-cobalt cathode families \textnormal{LIB\_NMC111}, \textnormal{LIB\_NMC523}, \textnormal{LIB\_NMC622}, and \textnormal{LIB\_NMC811}.

The mapping highlights consistently high aluminum requirements across technologies, strong nickel demand in wind turbine alloys and in battery cathodes alongside cobalt and manganese, heavy reliance on silicon in \textnormal{SPV\_cSi}, and the primary association of rare-earth elements with permanent-magnet direct-drive generators in wind turbines. A full list of material–component–product mappings and data sources is provided in Appendix~\ref{ap:1}, Table~\ref{tab:material_map_full}.

Materials are sourced from domestic production, imports, stock, and recovery from retired units. Production and import levels follow historical supercycle trends \cite{dvorkin2024understanding}. Initial stock is assumed to be zero and accumulates over time if unused. A conservative 10\% recovery rate is applied to retired wind, solar, and battery units, due to limited recycling infrastructure and data availability in the U.S.

Maryland’s access to national material supply is scaled to 1.6\%, based on its average share of U.S. GDP (1.9\%) and electricity consumption (1.3\%) in 2024. For energy sector allocation, each of the 14 materials is assigned a fixed share—either 10\% or 30\%—based on historical usage patterns in building new generation infrastructure, as reported in \cite{dvorkin2024understanding}. We acknowledge this simplification and highlight the need for further research on sectoral competition for materials.

\subsubsection{Lead Time for Deployment}

Lead time, defined as the delay from project initiation to operational status, accounts for both manufacturing and regulatory delays. Technology-specific values are adopted from \cite{decarolis2023annual}.

\subsubsection{Field Availability}

Generation expansion is limited by available land and offshore areas. Land is divided into technology-specific zones for LBW and SPV, and a shared common field accessible to LBW, SPV, and BSS. Land use is tracked dynamically as defined in Eqs.~\eqref{eq:field_status_all} and~\eqref{eq:initial_field}, with retired facilities returning their area to the common field for reuse. Field availability estimates follow \cite{moriarty2013feasibility}, and technology-specific capacity densities are based on \cite{mulas2023capacity, epa_repowering, offshorewindpowerhub, noland2022spatial}.

Technology-specific lead time, lifetime (see Section~\ref{GenerationandStorage}), and capacity density are summarized in Table~\ref{tab:leadtime_lifetime}.

\subsection{Generation and Storage Technologies}
\label{GenerationandStorage}

The SC-GEP model incorporates both existing and potential generation and storage resources. Existing units are based on the U.S. Energy Information Administration’s EIA-860 dataset \cite{eia860}. Hydroelectric generation, which contributes modestly to Maryland’s overall supply, is represented as a steady output based on historical capacity factors, reflecting its relatively stable production profile.

New capacity is limited to renewables and battery storage, consistent with Maryland’s regulation \cite{md_sb316_2025} and the current interconnection queue, which includes no new thermal projects. While thermal technologies are excluded, they can be reintroduced with minor model adjustments if needed. Ramping and startup/shutdown constraints are simplified, justified by the planned 2025 retirement of Brandon Shores and the flexibility of the remaining gas-dominated thermal fleet.

Each modeled technology is assigned a standardized abbreviation: land-based wind (LBW), offshore wind (OSW), solar photovoltaics (SPV), battery storage systems (BSS), natural gas combined-cycle (NGCC), natural gas combustion turbines (NGCT), nuclear power (NUC), hydroelectric (HYD), petroleum-fired (OIL), biomass (BIO), and coal-fired (COAL). 

Technology-specific lifetimes follow \cite{mirletz2024annual}. Existing units retain their original design lifetimes without further extension. Units that have exceeded their design life are assumed to retire in the second year, allowing one additional year of operation to reflect current regulatory or technical extensions.

\begin{table}[b]
\centering
\caption{Lead Time, Lifetime, and Capacity Density.}
\label{tab:leadtime_lifetime}
\renewcommand{\arraystretch}{1.2}
\resizebox{\linewidth}{!}{%
\begin{tabular}{| c | c | c | c | c | c |}
\hline
 & BIO & BSS & COAL & HYD & LBW \\
\hline
Lead Time (yr)       & --  & 1   & --   & --   & 3   \\
\hline
Lifetime (yr)        & 45  & 15  & 30   & 100  & 30  \\
\hline
Capacity Density (MW/km\textsuperscript{2}) 
                     & 500 & 900 & 5000 & 3.59 & 3.09 \\
\hline
\end{tabular}
}

\resizebox{\linewidth}{!}{%
\begin{tabular}{| c | c | c | c | c |}
\hline
 & NGCC(CT) & NUC & OSW & SPV \\
\hline
Lead Time (yr)       & --       & --  & 4   & 2 \\
\hline
Lifetime (yr)        & 30       & 60  & 30  & 30 \\
\hline
Capacity Density (MW/km\textsuperscript{2}) 
                     & 3574     & 4723 & 5.2 & 36 \\
\hline
\end{tabular}
}
\\[2pt]
\footnotesize\textit{Note: “--” indicates the technology is not modeled for capacity expansion.}
\end{table}

\subsection{Data Assumptions and System Inputs}

\begin{figure*}[!b]
    \centering
    \includegraphics[width=\linewidth]{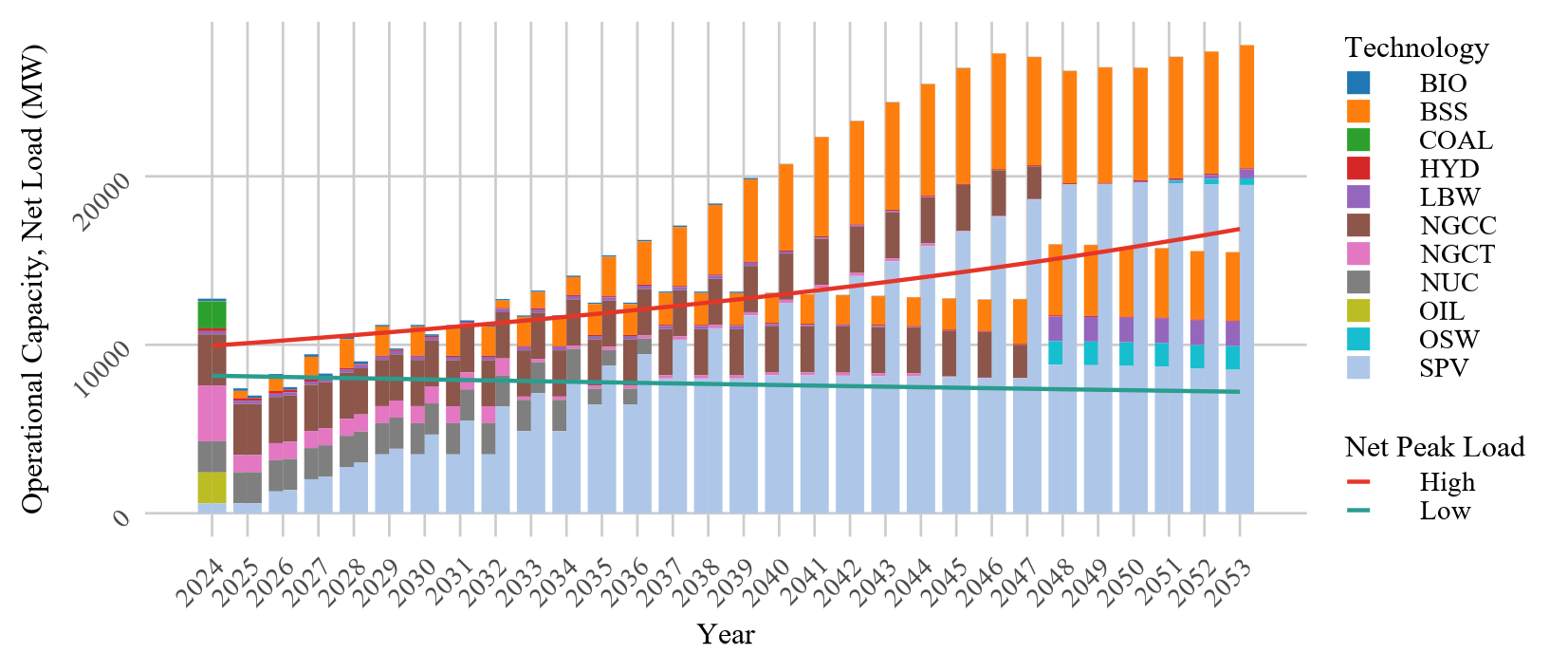}
    \caption{Operational capacity over the modeling horizon for baseline Low and High scenarios. Stacked bars represent technology-specific capacity, with Low baseline on the left in each year. Lines show net peak load, indicating system demand.}\label{fig:OperationalCapacity}
\end{figure*}

The planning horizon begins in 2024 and spans 30 years. Hourly load profiles are based on PJM data~\cite{pjm_dataminer2}, with demand growth projected using compound annual growth rates (CAGR) from Maryland’s Ten-Year Plan~\cite{mdpsc2023tenyear}. Imported power is modeled exogenously using PJM’s Hourly Net Exports by State~\cite{pjm_dataminer2} and allocated by peak load share. Two demand scenarios are considered: \textbf{Low}, which excludes aggressive electrification and data center growth, and \textbf{High}, which includes both. These trajectories, combined with the supply chain assumptions in Section~\ref{sec:supplychain}, define the baseline of the scenarios (\textit{baseline}). 

Each demand case is paired with two supply chain sub-scenarios.
The relaxed case (\textit{w/o SC}) removes all upstream supply chain constraints and assumes unlimited material availability, zero lead time, and expanded land and offshore areas (2× relative to the \textit{baseline}). This configuration reflects a traditional GEP formulation, in which technologies can be built at the quantities and times selected by the optimization,  without limits imposed by material supply, manufacturing capacity, or deployment conditions. The constrained case (\textit{lim. SC}) restricts materials to domestic and allied sources~\cite{dvorkin2024understanding}, incorporates manufacturing and deployment limits, and reflects rising geopolitical and trade-related risks.

\begin{table}[b]
\centering
\caption{Peak Load and CAGR Assumptions by Scenario.}
\label{tab:load_assumptions}
\resizebox{\linewidth}{!}{%
\LARGE
\begin{tabular}{|c|c|c|c|c|}
\hline
Zone & \multicolumn{2}{c|}{Low Scenario} & \multicolumn{2}{c|}{High Scenario} \\
\cline{2-5}
     & Peak Load (MW) & CAGR (\%) & Peak Load (MW) & CAGR (\%) \\
\hline
APS     & 1554 & 0.21 & 1683 & 4.67 \\
\hline
BGE     & 6428 & -0.65 & 6491 & 0.60 \\
\hline
DPL     &  961 & -0.45 & 1036 & 0.42 \\
\hline
PEPCO   & 2958 & 0.20 & 4472 & 0.65 \\
\hline
\end{tabular}}
\end{table}

Table~\ref{tab:load_assumptions} summarizes the peak loads and growth rates used across scenarios. 
A 15\% planning reserve margin is enforced, with ELCC values from PJM resource adequacy studies~\cite{pjm_planning}. Cost parameters are  from the 2024 U.S. capital cost benchmarks~\cite{us2024capital}, including investment, fixed, and variable O\&M costs. VOLL is set at \$10{,}000/MWh~\cite{hansen2024shortage}. The RPS violation penalty is \$60/MWh, based on Maryland’s alternative compliance payment, and the reserve margin shortfall penalty is \$263{,}000/MW-year, based on PJM’s Net Cost of New Entry (CONE) for 4-hour battery storage~\cite{pjm_planning}.

A complete list of case study assumptions is provided in the Appendix~\ref{A.2}, Table~\ref{tab:simulation_assumptions}, with all parameter sources documented therein.

\begin{figure*}[t]
    \centering
    \includegraphics[width=\linewidth]{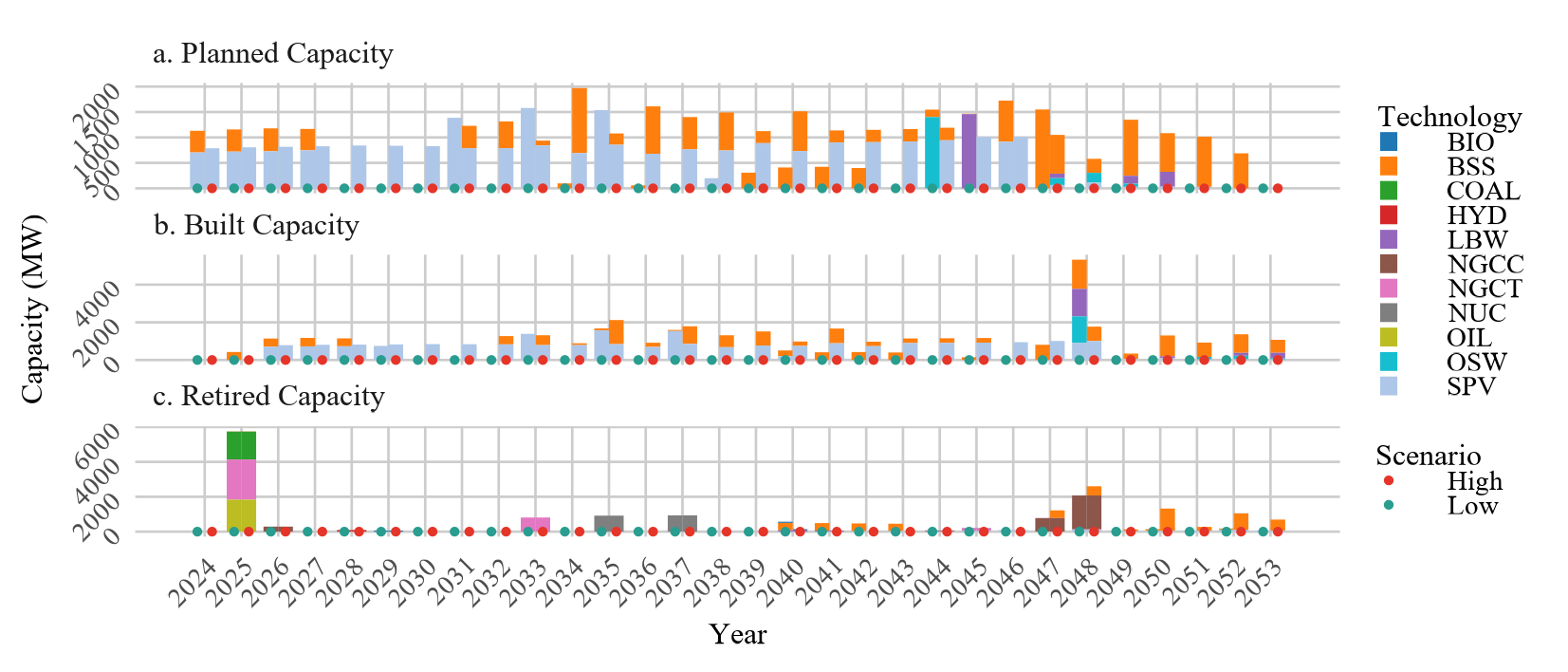}
    \caption{Optimization Results for Status Variables: (a) Planning Capacity, (b) Built Capacity, and (c) Retirement Capacity. Stacked bars represent technology-specific capacity, with Low baseline on the left in each year.}
    \label{fig:Status}
\end{figure*}

\begin{figure*}[t]
    \centering
    \includegraphics[width=\linewidth]{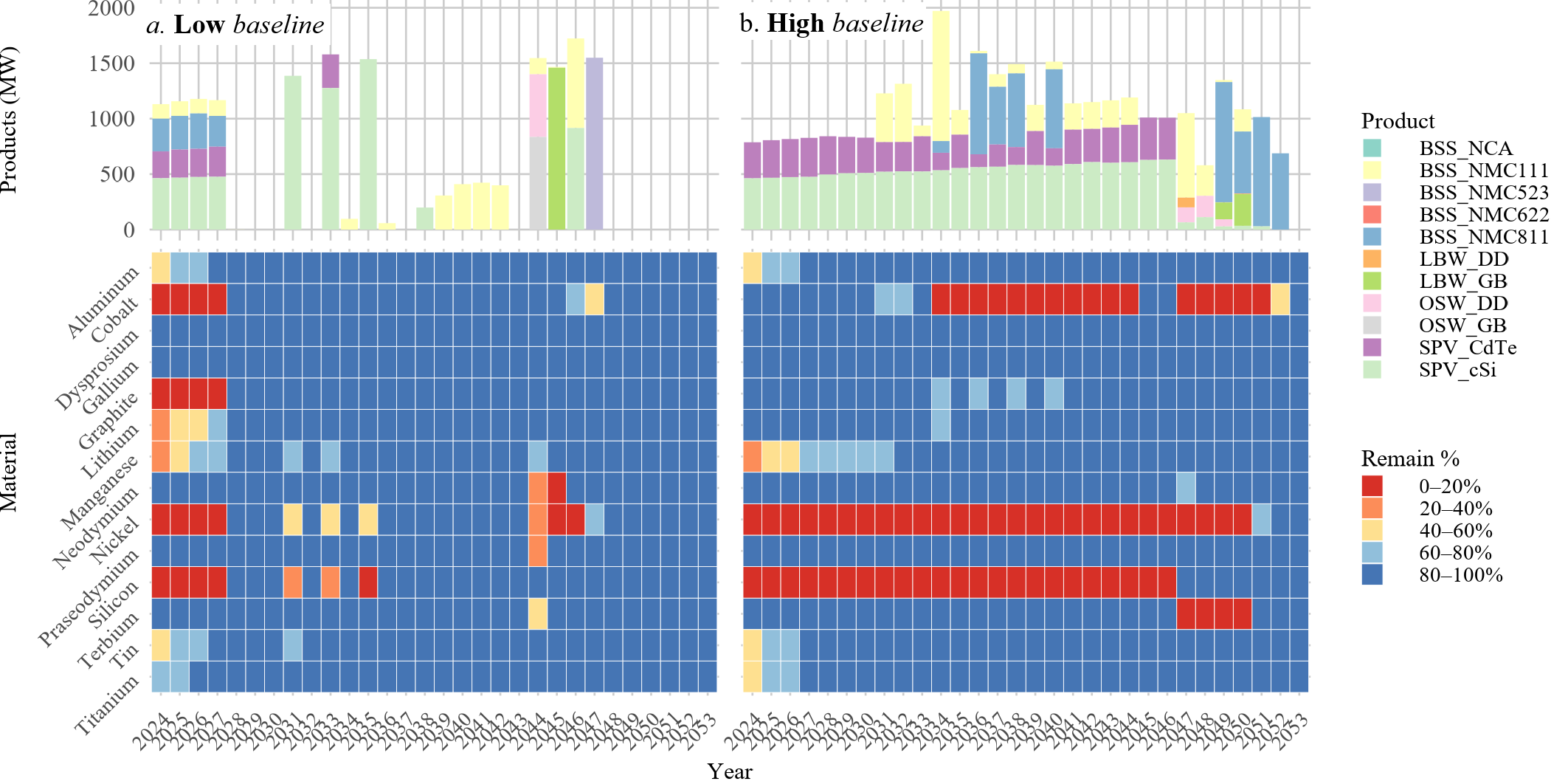}
    \caption{Planned product by technology (top) and remaining material availability (bottom) for baseline Low and High scenarios.}
    \label{fig:material_product}
\end{figure*}

\subsection{Numerical Results}

Figure~\ref{fig:OperationalCapacity} illustrates the optimized operational capacity changes from 2024 to 2053 under both the \textbf{Low} and \textbf{High} \textit{baseline} scenarios. The \textbf{Low} \textit{baseline} scenario exhibits greater diversification in generation technologies over time. By 2053, it achieves 4.1~GW of BSS, 1.5~GW of LBW, 1.4~GW of OSW, and 8.5~GW of SPV. In contrast, the \textbf{High} \textit{baseline} scenario reaches 7.3~GW of BSS, 0.5~GW of LBW, 0.4~GW of OSW, and 19.5~GW of SPV.

Figure~\ref{fig:Status} illustrates the dynamic interactions among SC-GEP status variables, e.g,. investment, construction, and retirement, under lead time constraints for both the \textbf{Low} and \textbf{High} \textit{baseline} scenarios. To ensure generation adequacy, capacity planning must anticipate upcoming retirements and initiate investments in advance. Several key retirement waves are observed: in 2025, driven by the forced retirement of units reaching their technical lifetimes at the start of the horizon; in 2033, marked by the retirement of Essential Power Rock Springs; in 2035 and 2037, corresponding to the sequential retirement of the two Calvert Cliffs nuclear units; and in 2047–2048, involving major NGCC plants, including CPV St. Charles, Wildcat Point, and Keys Energy Center. Capacity expansion decisions for generation and storage are  timed ahead of these retirements to preserve system adequacy.

Further insights from Figure~\ref{fig:material_product} show the planned product quantities by technology and the remaining material availability over time. Material consumption is assumed to occur during the planning stage (Figure~\ref{fig:Status}, a) and the planned products (Figure~\ref{fig:material_product}, top) contribute directly to the planned capacity, which precedes final plant commissioning (Figure~\ref{fig:Status}, b) due to construction lead times. Consequently, material consumption and product procurement are recorded years prior to the resulting operational capacity additions. Following the 2025 retirement shock, the system must quickly restore reliable capacity to maintain reserve margins and replace thermal units. Technologies with high ELCC, such as BSS, are favored for their contribution to peak reliability. In the \textbf{Low} \textit{baseline} scenario, the model deploys both BSS and SPV, with SPV supporting daytime peaks. Their short lead times enable a rapid system response. In contrast, the \textbf{High} \textit{baseline} scenario prioritizes SPV due to faster demand growth. SPV is preferred for its fast deployment, lower material intensity, and alignment with peak-hour loads to reduce VOLL penalties. However, the supply of bottleneck materials such as silicon and nickel remains insufficient to support further BSS deployment before 2031.

Material constraints strongly influence early technology choices before 2031. In both the \textbf{Low} and \textbf{High} \textit{baseline} scenarios, initial SPV deployment includes a mix of c-Si and CdTe products, reflecting silicon saturation and a shift toward CdTe to diversify supply. Nickel, needed for racking systems, also faces supply limitations. These bottlenecks restrict SPV deployment and introduce trade-offs: in the \textbf{Low} \textit{baseline} scenario, constrained materials must support both SPV and BSS to compensate for near-term capacity shortfalls, forcing the model to balance reliability needs against material availability.

After 2031, particularly in the \textbf{Low} \textit{baseline} scenario where load declines over time, new capacity is not planned unless triggered by major retirements. A notable shift occurs around 2044–2045, when LBW with gearbox designs becomes increasingly preferred over SPV. This transition is driven by cost dynamics: after 2045, the discounted adjusted capital cost of LBW (\$28k/MW/yr) falls below that of SPV (\$29k/MW/yr). Additionally, limited capacity needs between 2036 and 2043 allow constrained materials to accumulate, enabling the deployment of more material-intensive technologies. OSW is also planned during this period. In 2044, both gearbox-based and direct-drive OSW are deployed in anticipation of NGCC retirements in the DPL zone by 2048 and due to field constraints that limit further onshore expansion. Despite higher capital costs, OSW is needed to maintain reliability under spatial limitations. The model balances resource use between OSW types: gearbox OSW uses more nickel but less neodymium, while direct-drive OSW consumes significantly more neodymium, a rare earth element with severe supply constraints. Diversifying between the two enables more efficient use of limited materials while preserving availability for future needs. After 2049, no additional capacity is needed, allowing the remaining land to be used for more land-intensive LBW without limiting further deployment.

In the \textbf{High} \textit{baseline} scenario, continuous load growth places sustained pressure on meeting energy and reserve margin needs. To minimize costly unserved energy penalties, rapid generation capacity deployment is prioritized, driving ongoing SPV expansion through 2046. Both c-Si and CdTe SPV are used, depending on material constraints. As availability improves after 2031, the system also invests in NMC-based storage—favoring NMC 111 from 2031–2035 when cobalt is more available, and shifting to NMC 811 from 2036–2040 and 2049–2052 as nickel becomes more accessible. Wind deployment begins in 2047, despite LBW becoming more cost-effective than SPV by 2046. The delay reflects the urgency to meet rising demand, where generation and reserve margin shortfall penalties outweigh cost differences between technologies. Under material constraints, both c-Si and CdTe SPV continue expanding as long as they support resource adequacy. By 2047, SPV alone no longer suffices, prompting deployment of higher-capacity-factor technologies like LBW and OSW. Due to nickel limits, direct-drive OSW—using less nickel than gearbox designs—is preferred. In parallel, lower-nickel BSS options, especially NMC 111, are selected to maintain resource adequacy while easing material bottlenecks.

\begin{figure*}[t]
    \centering
    \includegraphics[width=1\linewidth]{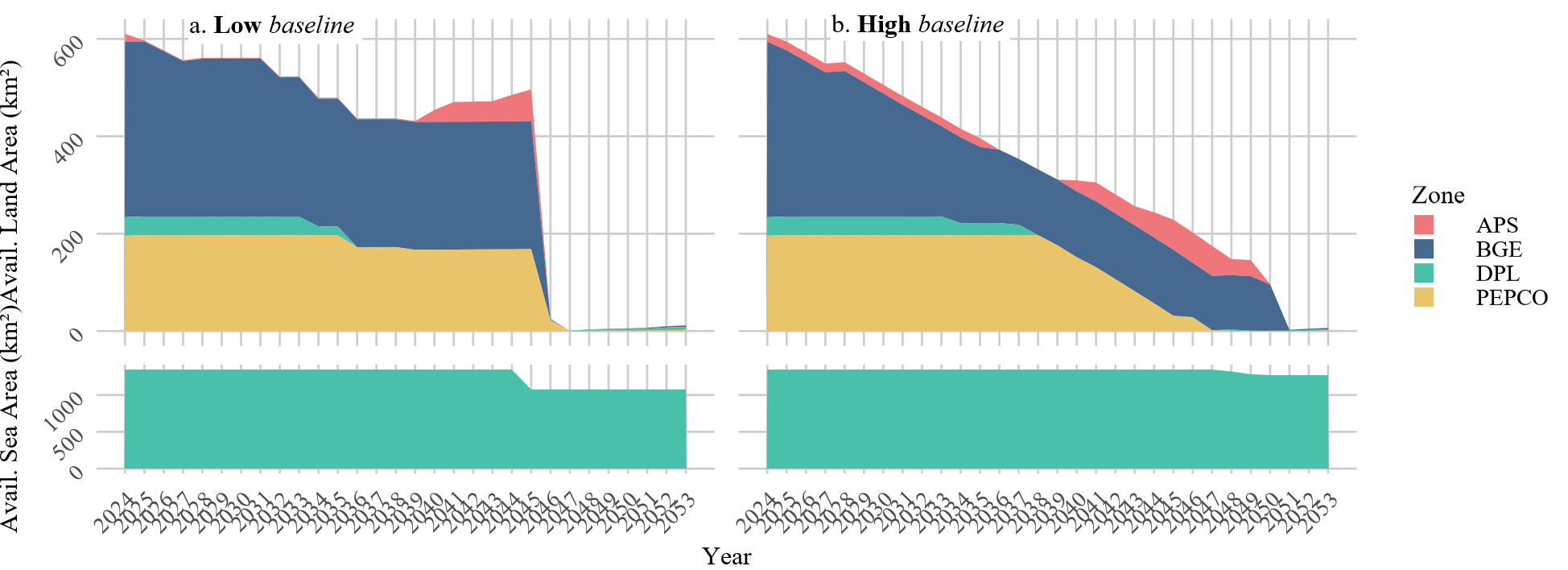}
    \caption{Field availability at the beginning of each year: land (top) and offshore (bottom) area for new deployment.}
    \label{fig:land_use}
\end{figure*}

\begin{figure*}[t]
    \centering
    \includegraphics[width=1\linewidth]{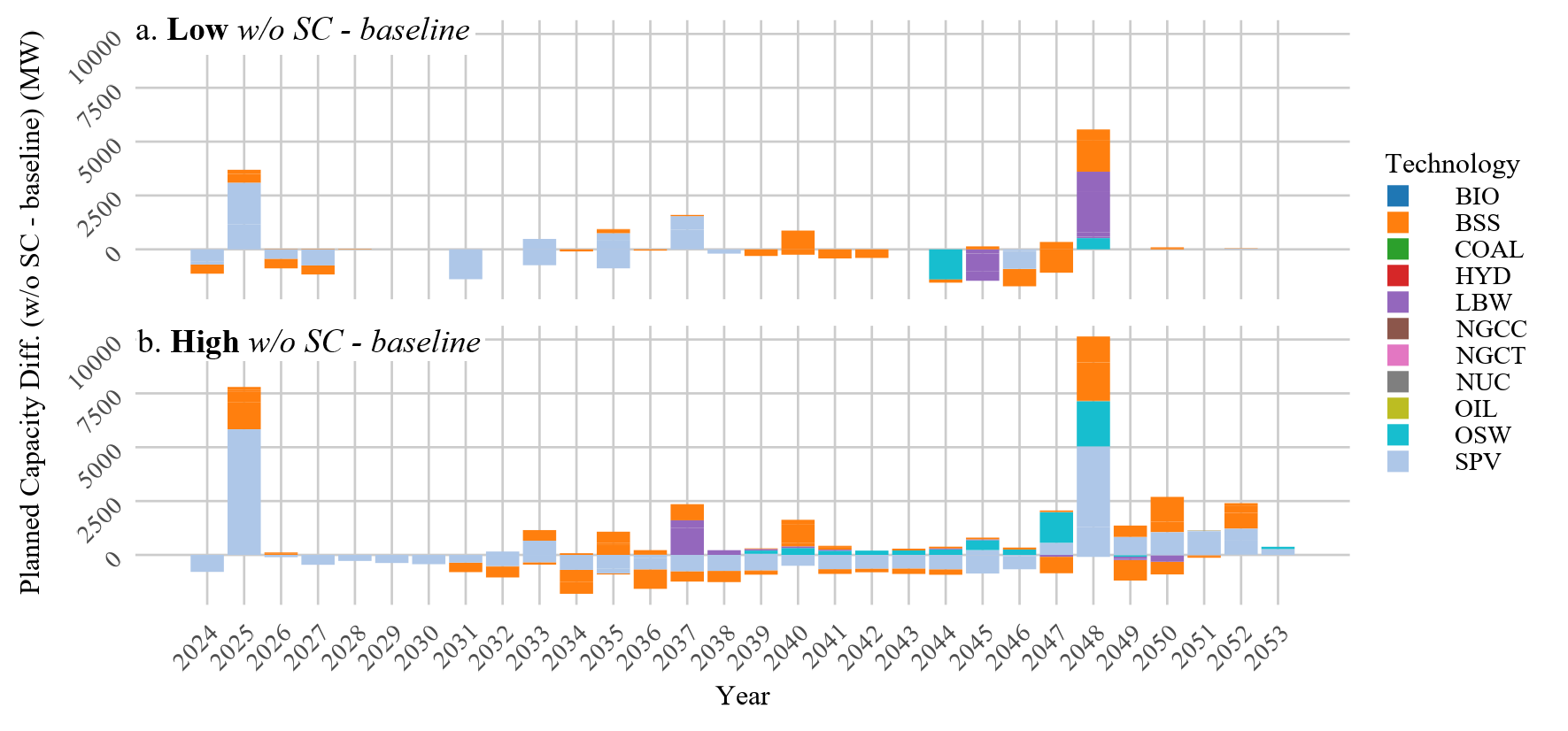}
    \caption{Difference in planned capacity by technology between the w/o SC and baseline sub-scenarios. Positive values indicate technologies favored in w/o SC.}
    \label{fig:plan_diff}
\end{figure*}

As the system moves into the early 2050s, land availability becomes a binding constraint, as illustrated in the right part of Figure~\ref{fig:land_use}, with most land outside the BGE service territory fully allocated after 2049. To sustain capacity expansion under these spatial limits, the model increasingly turns to BSS, which requires minimal land and offers short lead times. With nickel availability gradually improving and cobalt supplies tightening, the system strategically shifts toward a greater share of NMC 811, which uses less cobalt, while maintaining a smaller share of NMC 111 to balance the evolving material constraints.

\begin{figure*}[t]
    \centering
    \includegraphics[width=1\linewidth]{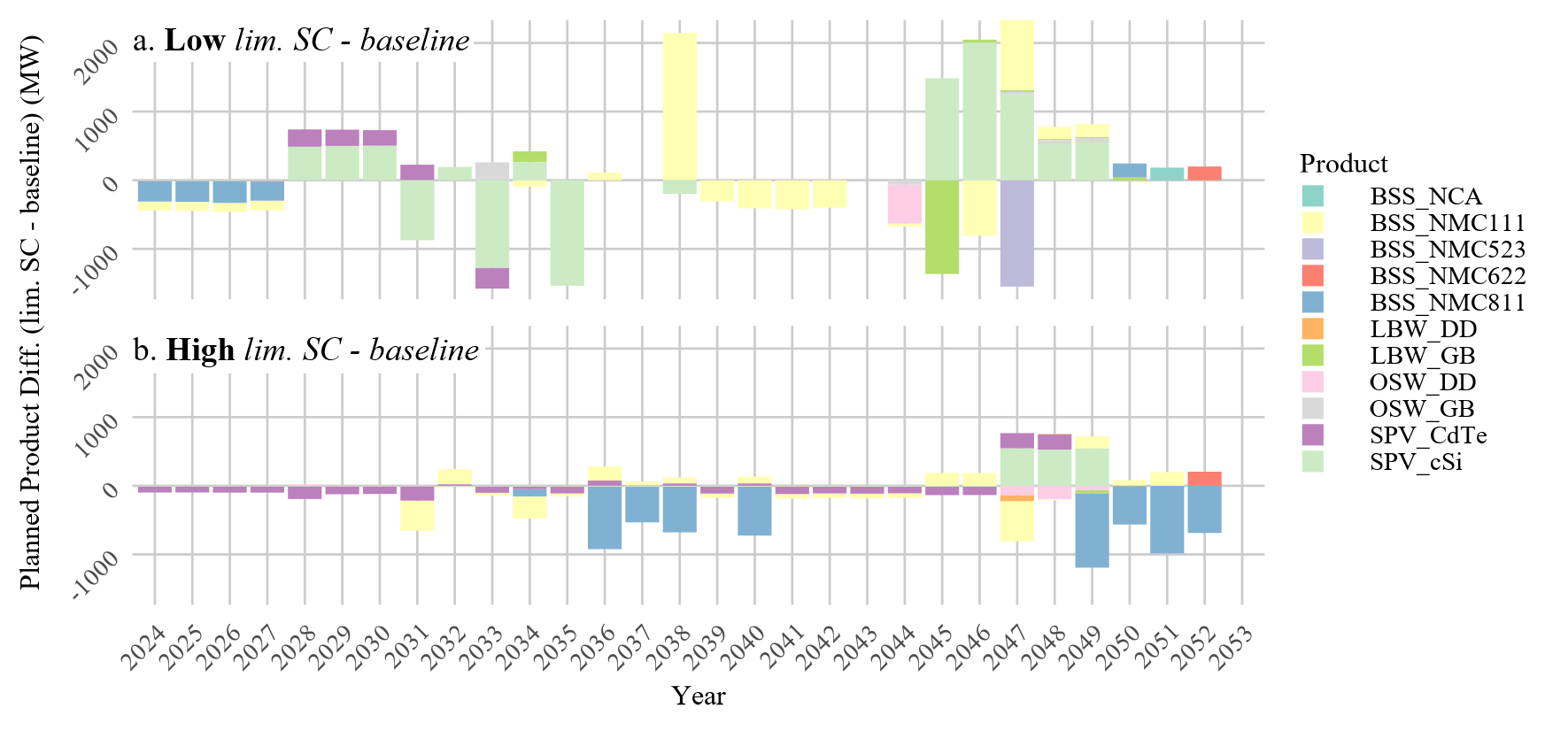}
    \caption{Difference in planned product deployment between the lim. SC and baseline sub-scenarios. Positive values indicate technologies favored in lim. SC.}
    \label{fig:prod_diff}
\end{figure*}

We compare the \textbf{Low} and \textbf{High} \textit{baseline} scenarios against two alternatives: \textit{w/o SC}, shown in Figure~\ref{fig:plan_diff}, and \textit{lim. SC}, shown in Figure~\ref{fig:prod_diff}. In the comparison with \textit{w/o SC}, the analysis focuses on planned capacity at the technology level, as this scenario assumes unlimited material availability and allows unrestricted product selection within each technology. In contrast, the comparison with \textit{lim. SC} emphasizes product-level choices, since stricter material constraints limit both the scale and the type of deployable technologies.

Both the \textbf{Low} and \textbf{High} \textit{w/o SC} scenarios show reactive, just-in-time planning enabled by the absence of lead time and material constraints. This flexibility allows immediate responses to major retirements. In the \textbf{Low} scenario, LBW is added in 2048 due to relaxed land limits and its cost advantage. In the \textbf{High} scenario, sustained load growth drives greater high-capacity-factor OSW deployment after 2045, unconstrained by material availability.

In both the \textbf{Low} and \textbf{High} \textit{lim. SC} scenarios, tighter constraints on critical materials, especially rare earth elements, limit the viability of LBW and OSW. After 2045, no additional wind capacity is deployed. In contrast, silicon remains relatively available through domestic production and stable imports from allied countries, making c-Si SPV a more viable alternative. As a result, limited resources like nickel are redirected toward SPV. BSS also shift in response to these material constraints: from nickel-intensive NMC 811 to NMC 111 batteries, which require less nickel. This shift is more significant in the \textbf{High} \textit{lim. SC} scenario, where greater system stress and higher load amplify the pressure to adopt less resource-intensive technologies.

\begin{figure*}[t]
    \centering
    \includegraphics[width=\linewidth]{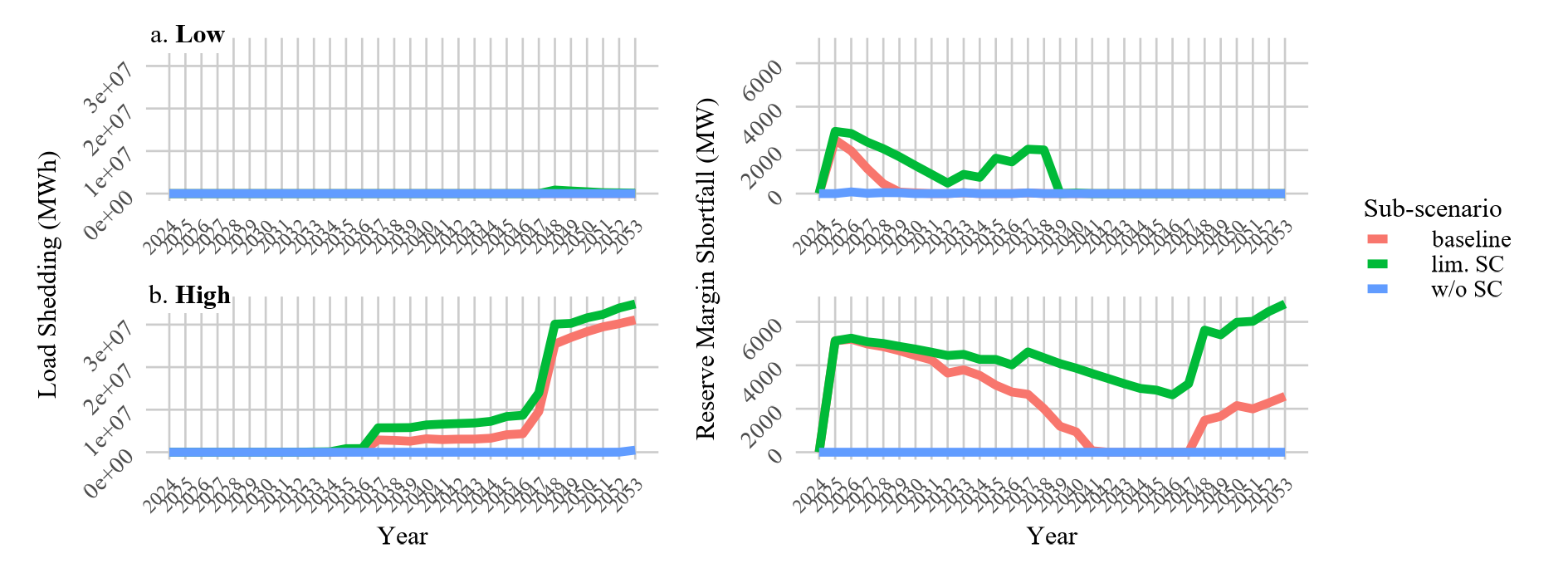}
    \caption{Yearly load shedding and reserve margin shortfalls across all scenarios: MWh load shedding (left) and MW reserve margin shortfall (right).}
    \label{fig:ls_rm}
\end{figure*}

Figure~\ref{fig:ls_rm} presents annual load shedding and reserve margin shortfalls (defined as the magnitude of capacity deficiency relative to the required reliability target) across all scenarios. Both the \textbf{Low} and \textbf{High} \textit{w/o SC} scenarios show virtually no reliability issues, as the system can respond immediately without supply chain constraints. In the \textbf{Low} \textit{baseline} scenario, load shedding is fully avoided, and reserve margin shortfalls are limited to early years due to the inability to replace 2025 forced retirements, constrained by lead times and material limits. In contrast, the \textit{lim. SC} scenario faces more severe reliability challenges: between 2033 and 2037, reserve margin shortfalls accumulate as material delays prevent timely replacement of retiring units, including Essential Power Rock Springs (2033) and Calvert Cliffs nuclear units (2035, 2037). In the \textbf{High} \textit{baseline} and \textit{lim. SC} scenarios, sustained load growth leads to load shedding from 2037 onward. Bottleneck materials including nickel, silicon, and cobalt delay timely capacity expansion, resulting in increasing shortages and reserve margin violations, which are caused by large-scale 2025 retirements  and are partially resolved by 2041 in the \textit{baseline} scenario. However, another significant gap (about 1.8 GW) emerges in 2048–2049 due to the retirement of 2 GW of NGCC capacity, and remains unresolved through 2053 as high mitigation costs under severe constraints discourage investment. In the \textit{lim. SC} scenario, reserve margin violations persist throughout the planning horizon.

\section{Conclusion}

This paper demonstrates that upstream supply chain constraints significantly shape generation expansion outcomes. In the Maryland case study, retiring capacity can be fully replaced by clean generation without reliability concerns assuming unlimited material availability, zero lead times, and ample land or offshore area. Under such ideal conditions, the total investment in the \textbf{Low} scenario is \$22.5 billion over the 30-year planning horizon from 2024 to 2053. When supply chain constraints are introduced, costs rise to \$23.7 billion, and reliability is compromised: a reserve margin shortfall of 2,471 MW emerges in 2025 following forced NGCC/NGCT retirements, persisting until 2029.

In the \textbf{High} scenario, the impact is more severe. Sustained load growth, coupled with material bottlenecks, leads to persistent reserve margin shortfalls that peak at 5,200 MW in 2024. Although temporarily mitigated by 2042, shortfalls resurface in 2047 due to a secondary wave of thermal retirements, remaining at 2,573 MW by 2053 and resulting in 342 GWh of load shedding. These results show that ignoring supply chain constraints not only underestimates system costs but also obscures significant risks to reliability. Given these limitations, just-in-time planning is no longer feasible, prompting the need for earlier and more strategic investments. The system must allocate limited resources such as materials, fields, and time to technologies (e.g., BSS and SPV) that can be deployed quickly to satisfy reserve margin requirements and reduce unserved energy penalties. These findings underscore the need to evaluate expansion plans under realistic supply chain conditions and suggest that lifetime extensions of existing assets may be necessary when deployment delays prevent timely additions.

Under more constrained conditions (\textit{lim. SC}), upstream limitations such as material scarcity and extended lead times override theoretical cost advantages, rendering technology diversification essential for system feasibility. Furthermore, these constraints introduce nonlinear dynamics where early, resource-intensive investments risk depleting critical materials required for future generation capacity. This increases the susceptibility to system over-correction \cite{ford2018simulating}, underscoring the critical necessity of anticipatory planning to mitigate material bottlenecks and ensure smooth energy transitions.

\section*{Acknowledgment}

The authors thank Dr. Ben Hobbs, Ziting Huang, and Stephanie Wilcox for their valuable assistance in the collection and preparation of data for the Maryland power system.

\appendices
\section{Material–Component–Product Mapping}\label{ap:1}
See Table~\ref{tab:material_map_full} for the complete material–component–product mapping and corresponding source references.

\begin{sidewaystable*}[htbp]
    \centering
    \scriptsize 
    \caption{Material–Component–Product Mapping with Data Sources}
    \label{tab:material_map_full}
    \renewcommand{\arraystretch}{1} 
    
    \begin{tabularx}{\textwidth}{l l l X p{2cm}}
        \toprule
        \textbf{Technology} & \textbf{Product} & \textbf{Component} & \textbf{Material (t/MW)} & \textbf{Source} \\
        \midrule
        
        \multirow{8}{*}{LBW} 
          & \multirow{4}{*}{LBW\_DD} 
            & LBW\_Array & Aluminum: 2.41 & \multirow{16}{*}{\cite{carrara2020raw, cooperman2023renewable, watari2019total, valero2018material, beylot2019mineral, tokimatsu2017energy, fishman2019impact, elshkaki2019energy, deetman2018scenarios, habib2016reviewing, cao2019resourcing, maanberger2018global, nassar2016byproduct, farina2022material}} \\
          & & LBW\_Foundation & Aluminum: 1.29e-3; Cobalt: 4.19e-4; Gallium: 7.88e-6; Lithium: 2.75e-6; Manganese: 3.74e-1; Nickel: 3.91e-1; Praseodymium: 1.01e-4; Titanium: 9.62e-3 & \\
          & & LBW\_Substation & Aluminum: 4.09e-3; Manganese: 1.39e-2; Nickel: 9.10e-3; Praseodymium: 2.70e-6; Titanium: 2.82e-4 & \\
          & & LBW\_Turbine\_DD & Aluminum: 5.81e-1; Cobalt: 2.44e-3; Dysprosium: 1.16e-2; Graphite: 4.60e-3; Manganese: 1.98; Nickel: 1.47; Neodymium: 1.99e-1; Praseodymium: 3.46e-3; Terbium: 7.64e-3 & \\
          \cmidrule{2-4}
          & \multirow{4}{*}{LBW\_GB} 
            & LBW\_Array & Aluminum: 2.41 & \\
          & & LBW\_Foundation & Aluminum: 1.29e-3; Cobalt: 4.19e-4; Gallium: 7.88e-6; Lithium: 2.75e-6; Manganese: 3.74e-1; Nickel: 3.91e-1; Praseodymium: 1.01e-4; Titanium: 9.62e-3 & \\
          & & LBW\_Substation & Aluminum: 4.09e-3; Manganese: 1.39e-2; Nickel: 9.10e-3; Praseodymium: 2.70e-6; Titanium: 2.82e-4 & \\
          & & LBW\_Turbine\_GB & Aluminum: 8.06e-1; Cobalt: 2.44e-3; Dysprosium: 1.89e-3; Gallium: 4.48e-5; Graphite: 4.60e-3; Lithium: 9.06e-4; Manganese: 1.98; Neodymium: 3.97e-2; Nickel: 2.12; Praseodymium: 5.33e-4; Titanium: 5.07e-2 & \\
        \cmidrule{1-4}

        \multirow{8}{*}{OSW} 
          & \multirow{4}{*}{OSW\_DD} 
            & OSW\_Array & Aluminum: 2.41e-1; Cobalt: 1.31e-5; Manganese: 1.11e-2; Nickel: 1.22e-2; Titanium: 3.10e-4  \\
          & & OSW\_Substation & Cobalt: 1.43e-4; Manganese: 1.35e-1; Nickel: 1.01e-1; Praseodymium: 3.61e-5; Titanium: 3.42e-3 & \\
          & & OSW\_Substructure & Cobalt: 3.18e-3; Gallium: 3.80e-5; Manganese: 3.70; Nickel: 2.77; Praseodymium: 1.00e-3; Tin: 2.55e-4; Titanium: 9.45e-2 & \\
          & & OSW\_Turbine\_DD & Aluminum: 7.62e-1; Cobalt: 1.33e-3; Dysprosium: 1.63e-2; Gallium: 1.59e-5; Graphite: 6.39e-3; Manganese: 1.06; Neodymium: 1.04e-1; Nickel: 8.62e-1; Praseodymium: 4.32e-2; Terbium: 3.74e-4; Tin: 2.79e-4; Titanium: 2.66e-2 & \\
          \cmidrule{2-4}
          & \multirow{4}{*}{OSW\_GB} 
            & OSW\_Array & Aluminum: 2.41e-1; Cobalt: 1.31e-5; Manganese: 1.11e-2; Nickel: 1.22e-2; Titanium: 3.10e-4 & \\
          & & OSW\_Substation & Cobalt: 1.43e-4; Manganese: 1.35e-1; Nickel: 1.01e-1; Praseodymium: 3.61e-5; Titanium: 3.42e-3 & \\
          & & OSW\_Substructure & Cobalt: 3.18e-3; Gallium: 3.80e-5; Manganese: 3.70; Nickel: 2.77; Praseodymium: 1.00e-3; Tin: 2.55e-4; Titanium: 9.45e-2 & \\
          & & OSW\_Turbine\_GB & Aluminum: 1.06; Cobalt: 1.33e-3; Dysprosium: 2.66e-3; Graphite: 6.39e-3; Lithium: 1.27e-3; Manganese: 1.06; Neodymium: 2.06e-2; Nickel: 1.24; Praseodymium: 6.64e-3; Terbium: 4.89e-5; Tin: 2.00e-4; Titanium: 2.66e-2 & \\
        \midrule

        \multirow{10}{*}{SPV} 
          & \multirow{5}{*}{SPV\_CdTe} 
            & SPV\_Inverter & Aluminum: 2.06e-1; Cobalt: 3.64e-5; Gallium: 4.36e-7; Manganese: 4.24e-2; Nickel: 3.18e-2; Silicon: 4.29e-6; Tin: 5.37e-5; Titanium: 1.08e-3 & \multirow{10}{=}{\cite{carrara2020raw, cooperman2023renewable, watari2019total, beylot2019mineral, tokimatsu2017energy, fizaine2015renewable, kavlak2015metal, moss2013potential, gervais2021raw, duan2016rethinking, zhou2020dynamic}} \\
          & & SPV\_Racking & Aluminum: 4.93e-1; Cobalt: 8.70e-4; Gallium: 1.20e-5; Graphite: 8.47e-4; Lithium: 9.53e-4; Manganese: 9.50e-1; Nickel: 7.73e-1; Praseodymium: 2.57e-4; Tin: 2.00e-4; Titanium: 2.43e-2 & \\
          & & SPV\_BOS & Cobalt: 8.48e-6; Manganese: 9.87e-3; Nickel: 7.40e-3; Praseodymium: 2.67e-6; Titanium: 2.52e-4 & \\
          & & SPV\_Transformer & Cobalt: 9.10e-6; Manganese: 1.06e-2; Nickel: 7.95e-3; Praseodymium: 2.87e-6; Titanium: 2.71e-4 & \\
          & & SPV\_Module\_CdTe & Aluminum: 9.05; Cobalt: 9.36e-6; Manganese: 5.02e-2; Nickel: 8.66e-3; Praseodymium: 2.06e-6; Silicon: 3.82e-2; Tin: 2.47e-3; Titanium: 1.55e-4 & \\
          \cmidrule{2-4}
          & \multirow{5}{*}{SPV\_cSi} 
            & SPV\_Inverter & Aluminum: 2.06e-1; Cobalt: 3.64e-5; Gallium: 4.36e-7; Manganese: 4.24e-2; Nickel: 3.18e-2; Silicon: 4.29e-6; Tin: 5.37e-5; Titanium: 1.08e-3 & \\
          & & SPV\_Racking & Aluminum: 4.93e-1; Cobalt: 8.70e-4; Gallium: 1.20e-5; Graphite: 8.47e-4; Lithium: 9.53e-4; Manganese: 9.50e-1; Nickel: 7.73e-1; Praseodymium: 2.57e-4; Tin: 2.00e-4; Titanium: 2.43e-2 & \\
          & & SPV\_BOS & Cobalt: 8.48e-6; Manganese: 9.87e-3; Nickel: 7.40e-3; Praseodymium: 2.67e-6; Titanium: 2.52e-4 & \\
          & & SPV\_Transformer & Cobalt: 9.10e-6; Manganese: 1.06e-2; Nickel: 7.95e-3; Praseodymium: 2.87e-6; Titanium: 2.71e-4 & \\
          & & SPV\_Module\_cSi & Aluminum: 1.06e1; Cobalt: 1.80e-6; Manganese: 5.76e-2; Praseodymium: 1.91e-6; Silicon: 2.94; Tin: 6.75e-2; Titanium: 1.54e-4 & \\
        \midrule

        \multirow{25}{*}{LIB} 
          & \multirow{5}{*}{LIB\_NCA} 
            & LIB\_Cases & Aluminum: 6.00e-2 & \multirow{25}{*}{\cite{shafique2022material, xu2020future, bongartz2021multidimensional, song2019material}} \\
          & & LIB\_Collector & Aluminum: 1.13e-1 & \\
          & & LIB\_Electrolyte & Lithium: 7.50e-3 & \\
          & & LIB\_Anode & Graphite: 2.30e-1 & \\
          & & LIB\_Cathode\_NCA & Aluminum: 7.50e-3; Lithium: 3.43e-2; Nickel: 1.99e-1; Cobalt: 3.94e-2 & \\
          \cmidrule{2-4}
          & \multirow{5}{*}{LIB\_NMC111} 
            & LIB\_Cases & Aluminum: 6.00e-2 & \\
          & & LIB\_Collector & Aluminum: 1.13e-1 & \\
          & & LIB\_Electrolyte & Lithium: 7.50e-3 & \\
          & & LIB\_Anode & Graphite: 2.30e-1 & \\
          & & LIB\_Cathode\_NMC111 & Aluminum: 3.00e-2; Cobalt: 9.95e-2; Lithium: 3.90e-2; Manganese: 9.19e-2; Nickel: 9.81e-2 & \\
          \cmidrule{2-4}
          & \multirow{5}{*}{LIB\_NMC811} 
            & LIB\_Cases & Aluminum: 6.00e-2 & \\
          & & LIB\_Collector & Aluminum: 1.13e-1 & \\
          & & LIB\_Electrolyte & Lithium: 7.50e-3 & \\
          & & LIB\_Anode & Graphite: 2.30e-1 & \\
          & & LIB\_Cathode\_NMC811 & Aluminum: 2.25e-2; Cobalt: 2.25e-2; Lithium: 2.75e-2; Manganese: 2.25e-2; Nickel: 1.88e-1 & \\
          \cmidrule{2-4}
          & \multirow{5}{*}{LIB\_NMC523} 
            & LIB\_Cases & Aluminum: 6.00e-2 & \\
          & & LIB\_Collector & Aluminum: 1.13e-1 & \\
          & & LIB\_Electrolyte & Lithium: 7.50e-3 & \\
          & & LIB\_Anode & Graphite: 2.30e-1 & \\
          & & LIB\_Cathode\_NMC523 & Aluminum: 5.75e-2; Cobalt: 5.75e-2; Lithium: 3.50e-2; Manganese: 8.75e-2; Nickel: 1.48e-1 & \\
          \cmidrule{2-4}
          & \multirow{5}{*}{LIB\_NMC622} 
            & LIB\_Cases & Aluminum: 6.00e-2 & \\
          & & LIB\_Collector & Aluminum: 1.13e-1 & \\
          & & LIB\_Electrolyte & Lithium: 7.50e-3 & \\
          & & LIB\_Anode & Graphite: 2.30e-1 & \\
          & & LIB\_Cathode\_NMC622 & Cobalt: 5.08e-2; Lithium: 3.25e-2; Manganese: 5.00e-2; Nickel: 1.58e-1 & \\
          
        \bottomrule
    \end{tabularx}
\end{sidewaystable*}

\section{Summary of Case Study Assumptions}
\label{A.2}

See Table~\ref{tab:simulation_assumptions} for the complete case study assumptions and their corresponding sources.

\begin{table*}[htbp]
    \centering
    \caption{Summary of Case Study Assumptions}
    \label{tab:simulation_assumptions}
    \small 
    \renewcommand{\arraystretch}{1} 
    
    \begin{tabularx}{\textwidth}{l l X c}
        \toprule
        Category & Item & Assumption Description & Source \\
        \midrule
        
        \multirow{6}{*}{Supply Chain} 
          & Material Flow & Includes 14 critical materials (overlap of USGS and DOE lists), mapping 26 components to 11 final products. Products within the same technology share common components (see Table~\ref{tab:material_map_full}). & \cite{applegate2023final, igogo2022america, dvorkin2024understanding} \\
          & Material Availability & Future availability projected from declared imports, domestic USGS mining output, and historical CAGR. & \cite{usgs2025mcs, dvorkin2024understanding} \\
          & Accessibility Scale & Fixed at 1.6\%, based on the average of Maryland's GDP share (1.9\%) and electricity consumption share (1.3\%) of the U.S. total. & \cite{EIA_MD_state_profile} \\
          & Lead Time & Technology-specific fixed lead times (see Table~\ref{tab:leadtime_lifetime}). & \cite{decarolis2023annual} \\
          & Land Availability & Initial availability for APS, BGE, DPL, PEPCO (km$^2$): Wind (0, 0, 0, 0), Solar (16, 64.1, 27.5, 50.4), Common (0, 296.3, 10.7, 145.3). & \cite{moriarty2013feasibility} \\
          & Sea Availability & Initial availability for DPL (1345.1 km$^2$); others are 0. & \cite{BOEM_MD_offshore_wind} \\
        \midrule

        \multirow{4}{*}{Generation \& Storage} 
          & Existing Capacity & Follows EIA 860 (2024 version). & \cite{eia860} \\
          & Candidate Capacity & Limited to LBW, OSW, SPV, and LIB, consistent with Maryland's climate, clean energy regulations, and the PJM interconnection queue. & \cite{md_sb316_2025} \\
          & Lifetime & Technology-specific lifetimes (see Table~\ref{tab:leadtime_lifetime}). No extension allowed. Units exceeding lifetime by 2024 receive 1-year grace before retirement. & \cite{mirletz2024annual} \\
          & Cost & Capital and variable costs follow 2024 U.S. capital cost benchmarks. & \cite{us2024capital} \\
        \midrule

        \multirow{5}{*}{System Operation} 
          & Spatial Topology & 4-node system (see Figure~\ref{fig:diagram}) based on utility service territory maps, and assume all PJM Window 3 transmission upgrades are constructed; no additional transmission expansion is modeled. & \cite{dvorkinenergy} \\
          & Load & Peak load and growth rates detailed in Table~\ref{tab:load_assumptions}. & \cite{pjm_dataminer2, mdpsc2023tenyear} \\
          & Reserve Margin & Modeled on ELCC basis (15\% Peak Load). Shortfall penalty: \$263,000/MW (based on PJM Net CONE for 4-hour battery). & \cite{pjm_planning} \\
          & VoLL Penalty & Set to \$10,000/MWh (based on PJM 2024 price report). & \cite{hansen2024shortage} \\
          & RPS & Mandates a 15\% Solar PV carve-out effective 2030. Violation penalty is \$60/MWh (based on Maryland's alternative compliance payment). & \cite{pjm_planning} \\

        \bottomrule
    \end{tabularx}
\end{table*}

\bibliographystyle{ieeetr}
\bibliography{bio}

\begin{thebibliography}{10}

\bibitem{atems2018effect}
B.~Atems and C.~Hotaling, ``The effect of renewable and nonrenewable electricity generation on economic growth,'' {\em Energy Policy}, vol.~112, pp.~111--118, 2018.

\bibitem{arora2016energy}
V.~Arora and S.~Shi, ``Energy consumption and economic growth in the united states,'' {\em Applied Economics}, vol.~48, no.~39, pp.~3763--3773, 2016.

\bibitem{galvin2022electric}
R.~Galvin, ``Are electric vehicles getting too big and heavy? modelling future vehicle journeying demand on a decarbonized us electricity grid,'' {\em Energy Policy}, vol.~161, p.~112746, 2022.

\bibitem{shehabi20242024}
A.~Shehabi, A.~Hubbard, A.~Newkirk, N.~Lei, M.~A.~B. Siddik, B.~Holecek, J.~Koomey, E.~Masanet, D.~Sartor, {\em et~al.}, ``2024 united states data center energy usage report,'' 2024.

\bibitem{doe2024}
{U.S. Department of Energy}, ``Evaluating the reliability and security of the united states electric grid,'' tech. rep., U.S. Department of Energy, July 2024.

\bibitem{schulze2024overcoming}
K.~Schulze, F.~Kullmann, J.~Weinand, and D.~Stolten, ``Overcoming the challenges of assessing the global raw material demand of future energy systems. joule 8, 1936-1957,'' 2024.

\bibitem{elshkaki2019energy}
A.~Elshkaki and L.~Shen, ``Energy-material nexus: The impacts of national and international energy scenarios on critical metals use in china up to 2050 and their global implications,'' {\em Energy}, vol.~180, pp.~903--917, 2019.

\bibitem{SEIA2022}
S.~E. I.~A. (SEIA), ``Trade and supply chain barriers delay impact of historic clean energy law,'' 2022.
\newblock Accessed: 2025-03-22.

\bibitem{WSJ2024}
T.~W.~S. Journal, ``Orsted replaces ceo as wind industry faces challenges,'' 2024.
\newblock Accessed: 2025-03-22.

\bibitem{Powell2024MorvenWindFarm}
E.~Powell, ``Bp's morven wind farm at risk of missing start date.'' \emph{The Times}, November 2024.
\newblock Accessed: 2025-03-28.

\bibitem{rong2012does}
F.~Rong and D.~G. Victor, ``What does it cost to build a power plant,'' {\em Laboratory on International Law and Regulation ILAR}, 2012.

\bibitem{iea2024coal}
{International Energy Agency}, ``Accelerating just transitions for the coal sector.'' International Energy Agency, 2024.
\newblock \url{https://www.iea.org/reports/accelerating-just-transitions-for-the-coal-sector}.

\bibitem{john2024supply}
L.~John, W.~D. Piotrowicz, and A.~Ruggiero, ``Supply chain disruptions and their impact on energy sector during covid-19,'' in {\em Sustainable and resilient supply chain}, vol.~12, pp.~65--92, Emerald Publishing Limited, 2024.

\bibitem{qiu2024impacts}
Y.~Qiu, G.~Iyer, N.~Graham, M.~Binsted, M.~Wise, P.~Patel, and B.~Yarlagadda, ``The impacts of material supply availability on a transitioning electric power sector,'' {\em Cell Reports Sustainability}, vol.~1, no.~10, 2024.

\bibitem{wilkerson2015comparison}
J.~T. Wilkerson, B.~D. Leibowicz, D.~D. Turner, and J.~P. Weyant, ``Comparison of integrated assessment models: carbon price impacts on us energy,'' {\em Energy Policy}, vol.~76, pp.~18--31, 2015.

\bibitem{pehnt2006dynamic}
M.~Pehnt, ``Dynamic life cycle assessment (lca) of renewable energy technologies,'' {\em Renewable energy}, vol.~31, no.~1, pp.~55--71, 2006.

\bibitem{zhang2023graphite}
J.~Zhang, C.~Liang, and J.~B. Dunn, ``Graphite flows in the us: insights into a key ingredient of energy transition,'' {\em Environmental Science \& Technology}, vol.~57, no.~8, pp.~3402--3414, 2023.

\bibitem{lara2018deterministic}
C.~L. Lara, D.~S. Mallapragada, D.~J. Papageorgiou, A.~Venkatesh, and I.~E. Grossmann, ``Deterministic electric power infrastructure planning: Mixed-integer programming model and nested decomposition algorithm,'' {\em European Journal of Operational Research}, vol.~271, no.~3, pp.~1037--1054, 2018.

\bibitem{zhang2022global}
L.~Zhang, Z.~Chen, C.~Yang, and Z.~Xu, ``Global supply risk assessment of the metals used in clean energy technologies,'' {\em Journal of Cleaner Production}, vol.~331, p.~129602, 2022.

\bibitem{patankar2023land}
N.~Patankar, X.~Sarkela-Basset, G.~Schivley, E.~Leslie, and J.~Jenkins, ``Land use trade-offs in decarbonization of electricity generation in the american west,'' {\em Energy and Climate Change}, vol.~4, p.~100107, 2023.

\bibitem{md_sb316_2025}
{Maryland General Assembly}, ``{Senate Bill 316: Climate and Energy Planning Act}.'' \url{https://mgaleg.maryland.gov/mgawebsite/Legislation/Details/sb0316?ys=2025RS}, 2025.
\newblock Maryland General Assembly, 2025 Regular Session.

\bibitem{7902215}
Y.~Liu, R.~Sioshansi, and A.~J. Conejo, ``Multistage stochastic investment planning with multiscale representation of uncertainties and decisions,'' {\em IEEE Transactions on Power Systems}, vol.~33, no.~1, pp.~781--791, 2018.

\bibitem{zou2019stochastic}
J.~Zou, S.~Ahmed, and X.~A. Sun, ``Stochastic dual dynamic integer programming,'' {\em Mathematical Programming}, vol.~175, pp.~461--502, 2019.

\bibitem{dvorkinenergy}
Y.~Dvorkin and B.~Hobbs, ``Energy resilience and efficiency in maryland,''

\bibitem{fazlollahi2014multi}
S.~Fazlollahi, S.~L. Bungener, P.~Mandel, G.~Becker, and F.~Mar{\'e}chal, ``Multi-objectives, multi-period optimization of district energy systems: I. selection of typical operating periods,'' {\em Computers \& Chemical Engineering}, vol.~65, pp.~54--66, 2014.

\bibitem{pjm2022rtep}
S.~Abdulsalam, ``2022 rtep window 3.'' \url{https://www.pjm.com/-/media/committees-groups/state-commissions/isac/2023/20231218/20231218-rtep-window-3-2022.ashx}, 2023.
\newblock Accessed: 2025-03-19.

\bibitem{short2011regional}
W.~Short, P.~Sullivan, T.~Mai, M.~Mowers, C.~Uriarte, N.~Blair, D.~Heimiller, and A.~Martinez, ``Regional energy deployment system (reeds),'' tech. rep., National Renewable Energy Lab.(NREL), Golden, CO (United States), 2011.

\bibitem{munoz2013approximations}
F.~D. Munoz, E.~E. Sauma, and B.~F. Hobbs, ``Approximations in power transmission planning: implications for the cost and performance of renewable portfolio standards,'' {\em Journal of Regulatory Economics}, vol.~43, pp.~305--338, 2013.

\bibitem{applegate2023final}
J.~D. Applegate, ``Final list of critical minerals,'' {\em US Geological Survey, Department of the Interior}, 2023.

\bibitem{igogo2022america}
T.~Igogo, ``America's strategy to secure the supply chain for a robust clean energy transition,'' tech. rep., USDOE Office of Policy, Washington DC (United States), 2022.

\bibitem{dvorkin2024understanding}
B.~Yao, H.~Jeong, M.~Mehrtash, B.~Allan, D.~Ockerman, and Y.~Dvorkin, ``Understanding supply chain constraints for the us clean energy transition,'' {\em npj Clean Energy}, vol.~1, no.~1, p.~9, 2025.

\bibitem{decarolis2023annual}
J.~DeCarolis and A.~LaRose, ``Annual energy outlook 2023,'' {\em US Energy Information Administration}, 2023.

\bibitem{moriarty2013feasibility}
K.~Moriarty, ``Feasibility study of anaerobic digestion of food waste in st. bernard, louisiana. a study prepared in partnership with the environmental protection agency for the re-powering america's land initiative: Siting renewable energy on potentially contaminated land and mine sites,'' tech. rep., National Renewable Energy Lab.(NREL), Golden, CO (United States), 2013.

\bibitem{mulas2023capacity}
D.~Mulas~Hernando, W.~Musial, P.~Duffy, and M.~Shields, ``Capacity density considerations for offshore wind plants in the united states,'' tech. rep., National Renewable Energy Laboratory (NREL), Golden, CO (United States), 2023.

\bibitem{epa_repowering}
{U.S. Environmental Protection Agency}, ``{RE-Powering America's Land},'' 2025.
\newblock Accessed: 2025-03-19.

\bibitem{offshorewindpowerhub}
{Offshore Wind Power Hub}, ``{Offshore Wind Power Hub},'' 2025.
\newblock Accessed: 2025-03-19.

\bibitem{noland2022spatial}
J.~K. N{\o}land, J.~Auxepaules, A.~Rousset, B.~Perney, and G.~Falletti, ``Spatial energy density of large-scale electricity generation from power sources worldwide,'' {\em Scientific Reports}, vol.~12, no.~1, p.~21280, 2022.

\bibitem{eia860}
{U.S. Energy Information Administration}, ``{Form EIA-860 Annual Electric Generator Report},'' 2025.
\newblock Accessed: 2025-03-25.

\bibitem{mirletz2024annual}
B.~Mirletz, L.~Vimmerstedt, G.~Avery, A.~Sekar, D.~Stright, D.~Akindipe, S.~Cohen, W.~Cole, P.~Duffy, A.~Eberle, {\em et~al.}, ``Annual technology baseline: The 2024 electricity update,'' tech. rep., National Renewable Energy Laboratory (NREL), Golden, CO (United States), 2024.

\bibitem{pjm_dataminer2}
{PJM Interconnection}, ``{Data Miner 2}.'' \url{https://dataminer2.pjm.com/}, 2023.
\newblock Accessed: 2025-03-19.

\bibitem{mdpsc2023tenyear}
{Maryland Public Service Commission}, ``Ten-year plan (2023–2032) of electric companies in maryland,'' tech. rep., Maryland Public Service Commission, 2023.
\newblock Accessed: 2025-03-19.

\bibitem{pjm_planning}
{PJM Interconnection}, ``{PJM Planning},'' 2025.
\newblock Accessed: 2025-03-19.

\bibitem{us2024capital}
U.~E.~I. Administration, ``Capital cost and performance characteristic estimates for utility-scale electric power generating technologies,'' 2024.

\bibitem{hansen2024shortage}
C.~Hansen, ``Recent shortage pricing efforts.'' \url{https://www.pjm.com/-/media/DotCom/committees-groups/task-forces/rcstf/2024/20241113/20241113-item-04---miso-shortage-pricing-update-to-pjm-rcstf.pdf}, Nov. 2024.
\newblock Presented at the PJM Reserve Certainty Senior Task Force meeting.

\bibitem{ford2018simulating}
A.~Ford, ``Simulating systems with fast and slow dynamics: lessons from the electric power industry,'' {\em System Dynamics Review}, vol.~34, no.~1-2, pp.~222--254, 2018.

\bibitem{carrara2020raw}
S.~Carrara, P.~Alves~Dias, B.~Plazzotta, and C.~Pavel, ``Raw materials demand for wind and solar pv technologies in the transition towards a decarbonised energy system,'' 2020.

\bibitem{cooperman2023renewable}
A.~Cooperman, A.~Eberle, D.~Hettinger, M.~Marquis, B.~Smith, R.~F. Tusing, and J.~Walzberg, ``Renewable energy materials properties database: Summary,'' tech. rep., National Renewable Energy Laboratory (NREL), Golden, CO (United States), 2023.

\bibitem{watari2019total}
T.~Watari, B.~C. McLellan, D.~Giurco, E.~Dominish, E.~Yamasue, and K.~Nansai, ``Total material requirement for the global energy transition to 2050: A focus on transport and electricity,'' {\em Resources, Conservation and Recycling}, vol.~148, pp.~91--103, 2019.

\bibitem{valero2018material}
A.~Valero, A.~Valero, G.~Calvo, and A.~Ortego, ``Material bottlenecks in the future development of green technologies,'' {\em Renewable and Sustainable Energy Reviews}, vol.~93, pp.~178--200, 2018.

\bibitem{beylot2019mineral}
A.~Beylot, D.~Guyonnet, S.~Muller, S.~Vaxelaire, and J.~Villeneuve, ``Mineral raw material requirements and associated climate-change impacts of the french energy transition by 2050,'' {\em Journal of Cleaner Production}, vol.~208, pp.~1198--1205, 2019.

\bibitem{tokimatsu2017energy}
K.~Tokimatsu, H.~Wachtmeister, B.~McLellan, S.~Davidsson, S.~Murakami, M.~H{\"o}{\"o}k, R.~Yasuoka, and M.~Nishio, ``Energy modeling approach to the global energy-mineral nexus: A first look at metal requirements and the 2 c target,'' {\em Applied energy}, vol.~207, pp.~494--509, 2017.

\bibitem{fishman2019impact}
T.~Fishman and T.~E. Graedel, ``Impact of the establishment of us offshore wind power on neodymium flows,'' {\em Nature Sustainability}, vol.~2, no.~4, pp.~332--338, 2019.

\bibitem{deetman2018scenarios}
S.~Deetman, S.~Pauliuk, D.~P. Van~Vuuren, E.~Van Der~Voet, and A.~Tukker, ``Scenarios for demand growth of metals in electricity generation technologies, cars, and electronic appliances,'' {\em Environmental science \& technology}, vol.~52, no.~8, pp.~4950--4959, 2018.

\bibitem{habib2016reviewing}
K.~Habib and H.~Wenzel, ``Reviewing resource criticality assessment from a dynamic and technology specific perspective--using the case of direct-drive wind turbines,'' {\em Journal of Cleaner Production}, vol.~112, pp.~3852--3863, 2016.

\bibitem{cao2019resourcing}
Z.~Cao, C.~O’sullivan, J.~Tan, P.~Kalvig, L.~Ciacci, W.~Chen, J.~Kim, and G.~Liu, ``Resourcing the fairytale country with wind power: a dynamic material flow analysis,'' {\em Environmental Science \& Technology}, vol.~53, no.~19, pp.~11313--11322, 2019.

\bibitem{maanberger2018global}
A.~M{\aa}nberger and B.~Stenqvist, ``Global metal flows in the renewable energy transition: Exploring the effects of substitutes, technological mix and development,'' {\em Energy Policy}, vol.~119, pp.~226--241, 2018.

\bibitem{nassar2016byproduct}
N.~T. Nassar, D.~R. Wilburn, and T.~G. Goonan, ``Byproduct metal requirements for us wind and solar photovoltaic electricity generation up to the year 2040 under various clean power plan scenarios,'' {\em Applied Energy}, vol.~183, pp.~1209--1226, 2016.

\bibitem{farina2022material}
A.~Farina and A.~Anctil, ``Material consumption and environmental impact of wind turbines in the usa and globally,'' {\em Resources, Conservation and Recycling}, vol.~176, p.~105938, 2022.

\bibitem{fizaine2015renewable}
F.~Fizaine and V.~Court, ``Renewable electricity producing technologies and metal depletion: A sensitivity analysis using the eroi,'' {\em Ecological Economics}, vol.~110, pp.~106--118, 2015.

\bibitem{kavlak2015metal}
G.~Kavlak, J.~McNerney, R.~L. Jaffe, and J.~E. Trancik, ``Metal production requirements for rapid photovoltaics deployment,'' {\em Energy \& Environmental Science}, vol.~8, no.~6, pp.~1651--1659, 2015.

\bibitem{moss2013potential}
R.~L. Moss, E.~Tzimas, H.~Kara, P.~Willis, and J.~Kooroshy, ``The potential risks from metals bottlenecks to the deployment of strategic energy technologies,'' {\em Energy policy}, vol.~55, pp.~556--564, 2013.

\bibitem{gervais2021raw}
E.~Gervais, S.~Shammugam, L.~Friedrich, and T.~Schlegl, ``Raw material needs for the large-scale deployment of photovoltaics--effects of innovation-driven roadmaps on material constraints until 2050,'' {\em Renewable and Sustainable Energy Reviews}, vol.~137, p.~110589, 2021.

\bibitem{duan2016rethinking}
H.~Duan, J.~Wang, L.~Liu, Q.~Huang, and J.~Li, ``Rethinking china's strategic mineral policy on indium: implication for the flat screens and photovoltaic industries,'' {\em Progress in Photovoltaics: Research and Applications}, vol.~24, no.~1, pp.~83--93, 2016.

\bibitem{zhou2020dynamic}
Y.~Zhou, J.~Li, H.~Rechberger, G.~Wang, S.~Chen, W.~Xing, and P.~Li, ``Dynamic criticality of by-products used in thin-film photovoltaic technologies by 2050,'' {\em Journal of cleaner production}, vol.~263, p.~121599, 2020.

\bibitem{shafique2022material}
M.~Shafique, M.~Rafiq, A.~Azam, and X.~Luo, ``Material flow analysis for end-of-life lithium-ion batteries from battery electric vehicles in the usa and china,'' {\em Resources, Conservation and Recycling}, vol.~178, p.~106061, 2022.

\bibitem{xu2020future}
C.~Xu, Q.~Dai, L.~Gaines, M.~Hu, A.~Tukker, and B.~Steubing, ``Future material demand for automotive lithium-based batteries,'' {\em Communications Materials}, vol.~1, no.~1, p.~99, 2020.

\bibitem{bongartz2021multidimensional}
L.~Bongartz, S.~Shammugam, E.~Gervais, and T.~Schlegl, ``Multidimensional criticality assessment of metal requirements for lithium-ion batteries in electric vehicles and stationary storage applications in germany by 2050,'' {\em Journal of cleaner production}, vol.~292, p.~126056, 2021.

\bibitem{song2019material}
J.~Song, W.~Yan, H.~Cao, Q.~Song, H.~Ding, Z.~Lv, Y.~Zhang, and Z.~Sun, ``Material flow analysis on critical raw materials of lithium-ion batteries in china,'' {\em Journal of Cleaner Production}, vol.~215, pp.~570--581, 2019.

\bibitem{usgs2025mcs}
{U.S. Geological Survey}, ``Mineral commodity summaries 2025,'' tech. rep., U.S. Geological Survey, 2025.
\newblock Version 1.2, March 2025, 212 p., ISSN 0076-8952 (print).

\bibitem{EIA_MD_state_profile}
{U.S. Energy Information Administration}, ``Maryland energy profile,'' 2025.

\bibitem{BOEM_MD_offshore_wind}
{Bureau of Ocean Energy Management}, ``Maryland offshore wind — state activities,'' 2025.

\end{thebibliography}

\end{document}